\documentclass[10pt,fleqn]{article}
\usepackage{amsmath,amssymb}
\usepackage{graphicx}
\usepackage{slashed}
\usepackage[margin=2.25cm]{geometry}
\usepackage[blocks]{authblk}
\usepackage[small]{caption}
\usepackage{subcaption}
\usepackage{cite}
\usepackage{multirow}
\usepackage{hhline}

\usepackage{dcolumn}
\usepackage{bm}
\usepackage{dsfont}
\usepackage{xcolor}
\usepackage{ulem}

\def\m{M_{\rm QCD}}

\def\aGG{\left\langle\alpha_s G^2 \right\rangle}
\def\mss{\left\langle m_s \overline{s}s\right\rangle}
\def\msqGq{\left\langle m_s\overline{q}\sigma G q\right\rangle}
\def\msqq{\left\langle m_s \overline{q}q\right\rangle}
\def\mqq{\left\langle m_q \overline{q}q\right\rangle}
\def\ss{\left\langle \bar s s\right\rangle}
\def\qq{\left\langle \bar q q\right\rangle}
\def\qGq{\left\langle \overline{q}\sigma G q\right\rangle}
\def\sGs{\left\langle \overline{s}\sigma G s\right\rangle}
\def\trho{\boldsymbol{\varrho}}

\begin{document}

\title{\sf 
Extending the Bridge Connecting Chiral Lagrangians and QCD Gaussian Sum-Rules for Low-Energy Hadronic Physics  
}

\author{Amir H. Fariborz\thanks{{fariboa@sunypoly.edu}}}
\affil{Department of Mathematics/Physics,  SUNY Polytechnic Institute, Utica, NY 13502, U.S.A.
}
\author{J. Ho\thanks{jason.ho@dordt.edu}}
\affil{%
	Department of Physics, Dordt University, Sioux Center, Iowa, 51250, USA
}%
\author{T.G. Steele\thanks{{Tom.Steele@usask.ca}}}
\affil{%
	Department of Physics and
	Engineering Physics, University of Saskatchewan, Saskatoon, SK,
	S7N 5E2, Canada
}%
\maketitle

\begin{abstract}
It has previously been demonstrated that   the mesonic fields in chiral Lagrangians can be related to the quark-level operators of QCD sum-rules via energy-independent (constant)  
scale factor matrices constrained by chiral symmetry. This leads   to universal scale factors for each type of  chiral nonet  related to  
quark-antiquark ($q\bar q$) operators and 
four-quark ($qq\bar q\bar q$) operators.
Motivated by these successful demonstrations of scale-factor universality for the  $K_0^*$ isodoublet and $a_0$  isotriplet scalar mesons, a revised  Gaussian QCD  sum-rule methodology is  developed that enables the extension to higher-dimensional isospin sectors, including the possibility of mixing with glueball components. Moreover, to extract non-perturbative information about a resonance stemming from the final state interactions of its
decay products, a background-resonance interference approximation is developed and shown to provide an excellent description of both 
$\pi K$ scattering amplitude data and $\pi\eta$ scattering calculations. 
This background-resonance interference approximation inspires new resonance models as ingredients in the scale-factor analysis connecting chiral Lagrangians and QCD Gaussian sum-rules.   
Using the revised Gaussian QCD sum-rule methodology, key properties of the scale factors are examined for the $K_0^*$ isodoublet and $a_0$  isotriplet scalar mesons  for a sequence  of increasingly sophisticated resonance models. 
Gaussian sum-rules are demonstrated to have sufficient resolution to distinguish between different resonance models, and it  is shown that the  background-resonance interference approximation  not only describes $\{\pi K,\pi\eta\}$ scattering, but leads to the best universality and energy-independence properties of the scale factors.      
\end{abstract}

\section{Introduction}

Although a complete framework for low-energy QCD is still missing, efforts to understand this challenging range of energy (see e.g., 
Refs.~\cite{PDG,PKZ_review}) continue on several fronts, including lattice QCD (see e.g., Refs.~\cite{LQCD1,LQCD2}), chiral perturbation theory (see e.g., Refs.~\cite{ChPT1,ChUA1,ChUA8,IAM1,IAM2,IAM3}),  QCD sum-rules 
\cite{SVZ} (see e.g., Refs.~\cite{Reinders:1984sr,Narison:2002woh,Gubler:2018ctz,Colangelo:2000dp} for reviews), different formulations of chiral Lagrangians \cite{Giacosa:2023fdz,Vereijken:2023jor,Jafarzade:2022uqo,GLSM_pieta,GLSM_piK,Parganlija:2012fy,GLSM_pipi,GLSM,GLSM_inst,Giacosa:2006tf,00_BFMNS,00_BFS61,NLCL_kappa,Carter:1995zi,Ko:1994en},  and various other approaches that aim to describe different aspects of low-energy processes (see e.g., Refs.~\cite{PDG,PKZ_review}).
QCD sum-rules provide a framework in which the resonances are probed through quark-level correlation functions calculated in QCD and then matched with the corresponding hadronic correlators 
through different forms of quark-hadron duality relations (e.g., integration with an exponential factor in Laplace sum-rules). While QCD sum-rule methodology has been quite successful, two aspects of this framework can be significantly enhanced:  (a) For resonances with substructures that are more complicated than a pure quark-antiquark  involving,  for example, mixing of quark-antiquark components with four-quark structures (i.e., structures with two quark fields and two antiquark fields),  and in the case of isosinglet scalar resonances also mixing with glueballs, the QCD operators used to probe these resonances should reflect their mixed content; (b) To simplify the analysis in QCD sum-rules, the hadronic side is often modeled by a narrow resonance approximation (or a Breit-Wigner shape) which is not expected to be accurate when probing resonances that are broad and interfering with nearby states.  In Refs.~\cite{CLQCDSR_2016,CLQCDSR:2019_Proc,CLQCDSR_2020,Fariborz:2019zht}, it was shown how point (a) can be addressed by a universal scale-factor connection of the QCD sum-rule framework to chiral Lagrangians (specifically the generalized linear sigma model (GLSM) of Ref.\cite{GLSM}).  In the present  work,  point (b) will be addressed by modeling the resonance shape for inclusion within the same scale-factor framework of Refs.~\cite{CLQCDSR_2016,CLQCDSR:2019_Proc,CLQCDSR_2020,Fariborz:2019zht}. Inspired by chiral Lagrangians, we introduce a new phenomenological approach to model the resonance shape that includes, for broad resonances,  the interaction of the resonance with the background (consisting of the final state interactions of the resonance decay products).  This new background-resonance interference approximation resonance shape can be recast as the sum of a pure Breit-Wigner part and a correction part that models the final state interactions of the main decay products.  The correction is determined by probing the contribution of the resonance to the appropriate scattering amplitude, together with the unitarity  of the scattering matrix.
	
Specifically,  in Refs.~\cite{CLQCDSR_2016,CLQCDSR:2019_Proc,CLQCDSR_2020,Fariborz:2019zht} we explored an interplay of chiral Lagrangians and QCD sum-rules based on the assumption that the mesonic building blocks and their corresponding quark composite currents are related to each other by scale factor matrices constrained by chiral symmetry to contain two universal and energy-independent (constant) scale factors. We examined the implication of this assumption in the isodoublet kappa channel \cite{CLQCDSR_2016} as well as in the isotriplet $a_0$ channel \cite{CLQCDSR_2020} and showed that, to a good approximation, the determined scale factors are nearly identical, supporting both the energy-independence constancy as well as the universality of the scale factors in these two systems.   At the level of chiral Lagrangians (for example, the generalized linear sigma model of Ref.~\cite{GLSM} which was applied in the analysis of Refs~\cite{CLQCDSR_2016,CLQCDSR:2019_Proc,CLQCDSR_2020,Fariborz:2019zht}), experimental data is used to determine the model parameters, and thereby the composition of the physical states is determined as a linear combination of their mesonic building blocks (composed of quark-antiquarks as well as four-quark 
structures constructed with two quark fields and two-antiquark fields).  The connection between chiral Lagrangians and QCD sum-rules introduced in Refs.~\cite{CLQCDSR_2016,CLQCDSR:2019_Proc,CLQCDSR_2020,Fariborz:2019zht} allows mirroring this linear combination at the composite quark level.  This allows us to use QCD sum rules to probe resonances via their admixtures of quark-antiquarks and four-quarks as determined by the chiral Lagrangians.

The GLSM of Ref.~\cite{GLSM} has also been applied to various processes such as $\pi\pi$ \cite{GLSM_pipi}, $\pi K$ \cite{GLSM_piK} and $\pi \eta$ \cite{GLSM_pieta} scattering. In these processes, the scattering amplitude is constructed and the
experimental data is analyzed. Scalar mesons play a particularly important role in these processes; for example, in $\pi K$ scattering the $K_0^*(700)$ kappa meson (and the heavier $K_0^*(1430)$ state) appear as poles in the scattering amplitude.  In the present work,, we develop and use a background-resonance interference approximation to extract the information about the broad kappa meson [and $K_0^*(1430)$] from the analysis of $\pi K$ scattering \cite{GLSM_piK} and incorporate that into the QCD sum-rule analysis.  
Since the kappa meson has a broad width and decays 100\% to $\pi K$ \cite{PDG}, the nonperturbative information about this state is hidden in the final state interactions of $\pi K$ and this information should be imported into the framework of QCD sum-rules applied to the kappa channel. As in this channel our focus is on the two resonances kappa and $K_0^*(1430)$,  which appear as poles in the $I=1/2$  $\pi K$ scattering amplitude, we approximate the scattering amplitude by a constant (complex) background and two $s$-channel poles with complex constants. This approximation to the scattering amplitude is then fitted to experimental data to determine these three constants. These constants should, presumably,  approximate the interaction of each resonance with the background as well as the interference of the two resonances with each other. We find  that this simple approximation provides a remarkable fit to experiment.  This background-resonance interference approximation results in a nontrivial resonance shape for the kappa meson and improvement in the energy-independence and universality scale factor properties in the sum-rule analysis. The same background-resonance interference approximation method is also applied to the analysis of isotriplets [$a_0(980)$ and $a_0(1450)$] which are probed in $\pi \eta$ scattering.   However, in this case there are no experimental data on $\pi\eta$ scattering amplitude, and therefore our study will be based on a self-consistent agreement with theoretical calculations within this framework.   
As in the isodoublet case, the background-resonance interference approximation model leads to improvement in the energy-independence and universality of scale factor properties.

The analysis methodology used in Refs.~\cite{CLQCDSR_2016,CLQCDSR:2019_Proc,CLQCDSR_2020,Fariborz:2019zht} to connect Chiral Lagrangians and QCD sum-rules becomes progressively more difficult to apply in higher-dimensional isospin sub-sectors of the scalar nonet, particularly with extensions to include mixing with gluonium content.
With the goal of extending our scale-factor framework to higher-dimensional systems, we develop a new sum-rule analysis methodology to address the limitations associated with higher-dimensional systems. This new Gaussian QCD sum-rule methodology is first compared to our previous scale factor analysis.  We then perform a systematic examination of different resonance models within our Gaussian QCD sum-rule analysis, focusing on the effect of the resonance model on the scale factors. We determine that the Gaussian QCD sum-rule scale-factor analysis is capable of discriminating between the resonance models, and find that the scale-factor properties contain supporting evidence for the resonance models inspired by the background-resonance interference approximation that describes $\{\pi K ,\pi\eta\}$ scattering. 

In Section~\ref{res_shape_section} we apply the generalized linear sigma model framework applied to $\pi K$ channel \cite{GLSM_piK} to develop the background-resonance interference approximation that is first fitted to $\pi K$ scattering data  and then fitted to $\pi\eta$ scattering calculations.  The background-resonance interference approximation is found to provide an excellent description to both channels, and provides inspiration for new resonance models in QCD sum-rules.  
The scale-factor methodology connecting Chiral Lagrangian mesonic fields to QCD quark-level two-quark and four-quark composite operators is reviewed in Section~\ref{scale_factor_section}. 
The revised scale factor methodology is developed in Section~\ref{new_scale_methodology_sec} and the Gaussian sum-rule relationships are developed to connect the scale factors, QCD, Chiral Lagrangians, and resonance properties.  
Section~\ref{models_section} presents a sequence of increasingly sophisticated resonance models, culminating with those inspired by the  background-resonance interference approximation.
Section~\ref{models_section} also examines the ability of Gaussian sum-rules to distinguish between different resonance models.
Results of the Gaussian QCD sum-rule scale-factor analysis is presented in Section~\ref{analysis_section} for different models in the $K_0^*$ isodoublet and $a_0$  isotriplet sectors, with a focus on the key properties of energy-independence and universality of the scale factors.  A summary and discussion of results is given in Section~\ref{summary_section}. 

\section{Chiral Lagrangian modeling of a resonance shape}
\label{res_shape_section}

The narrow resonance model for the  contribution of resonances to the hadronic correlator is only an approximation that requires  improvement when broader states are probed in QCD sum-rules.  The decay channels of the states and the rescattering effects of the decay products can provide an insight into the resonance shape.  In this section we study the kappa system and the $a_0$ system  probed in $\pi K$ and $\pi\eta$ scattering, respectively, and through the intricate  interaction of these states with the background (which includes the rescattering effects of the decay products as well as the underlying mixing of these states),  we develop a background-resonance interference approximation to extract the resonance shape.   

\subsection{The kappa system}
\label{kappa_subsection}

\begin{figure}[htb]
	\centering
    \includegraphics[width=0.4\columnwidth]{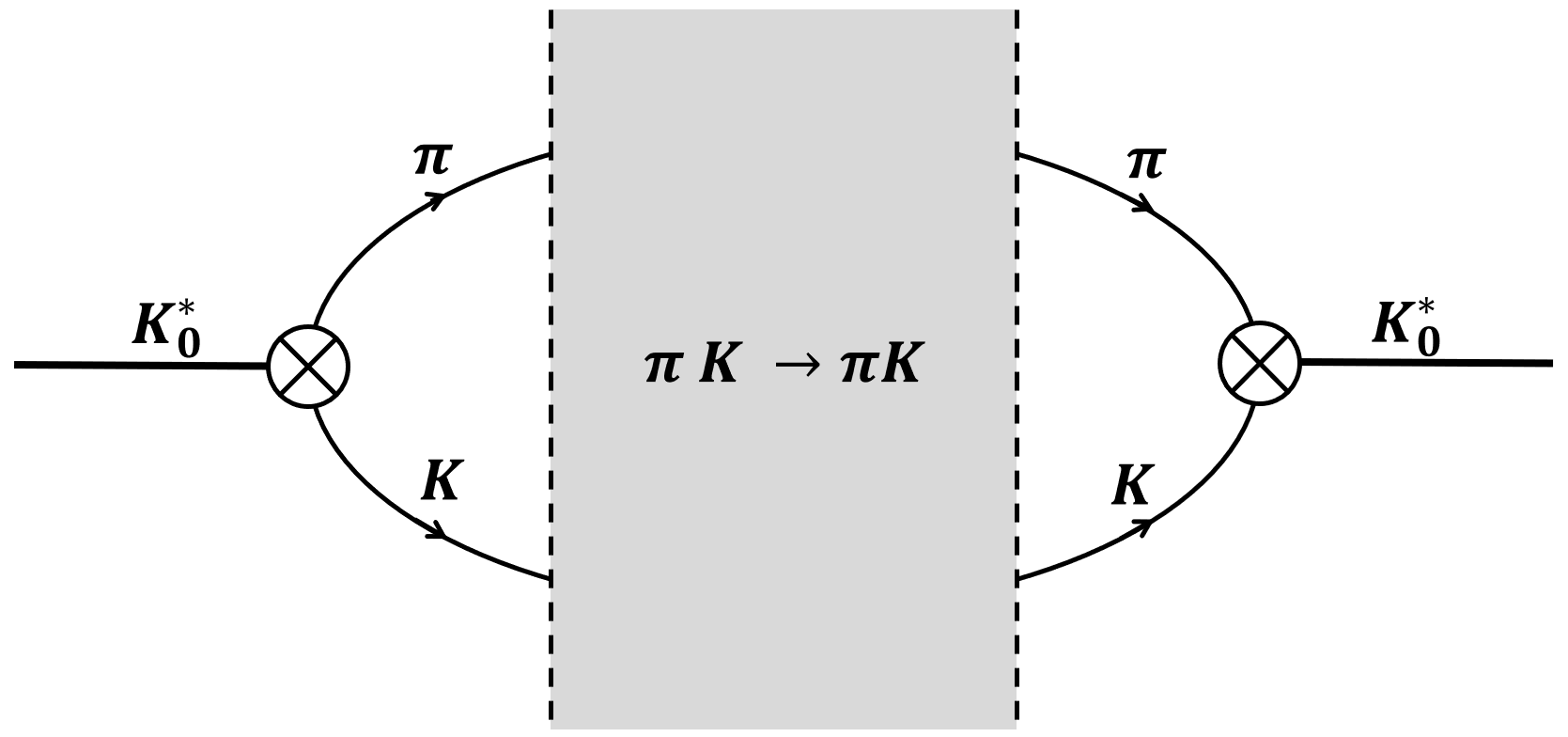}\hspace{0.02\columnwidth}
	\caption{Inclusion of $\pi K$ intermediate states in the kappa ($K_0^*$) correlator. }	\label{F_kappa_correlator}
\end{figure}

\begin{figure}[htb]
	\centering
	\includegraphics[width=0.4\columnwidth]{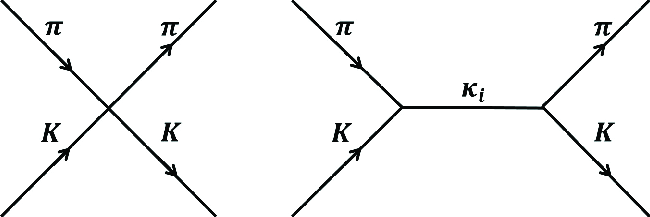}\hspace{0.02\columnwidth}
	\vskip 1cm
	\includegraphics[width=0.4\columnwidth]{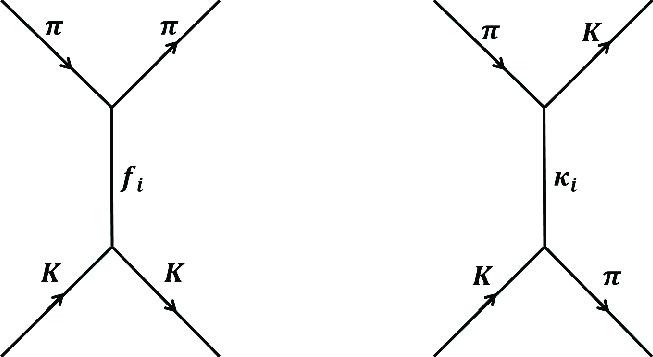}\hspace{0.02\columnwidth}
	\caption{Tree level Feynman diagrams for $\pi K$ scattering amplitude in the generalized linear sigma model.
	}
	\label{F_Feyn_diag}
\end{figure}

One of the broad scalar mesons below 1 GeV is the kappa mesons studied in different approaches \cite
{Pelaez_kappa,Oh:2006in,BES:2005frs,
Oller:2003vf,Jamin:2000wn}. Since the kappa meson strongly couples to the $\pi K$ channel, its contribution to the hadronic correlator should include the $\pi K$ rescattering effects
as shown in Fig.~\ref{F_kappa_correlator}.  
We take a phenomenological approach and probe the contribution of kappa and $K_0^*(1430)$ to the hadronic correlator  by connecting to their important roles in the $\pi K$ scattering amplitude.  
The full scattering amplitude is studied within the framework of the generalized linear sigma model in \cite{GLSM_piK} and includes contributions from $s$, $t$ and $u$ channels as well as a direct contact interaction shown in Fig.~\ref{F_Feyn_diag}. The $s$ and $u$ channels involve $\kappa$ [$K_0^*(700)$] and $K_0^*(1430)$ while the $t$-channel involves the exchanges of isosinglet states $f_i$.  

As pointed out in \cite{GLSM_piK} the unitarized scattering amplitude can be approximated  by the two poles and a background.  This modeling of the scattering amplitude is particularly helpful to the present study because it directly probes the way in which the propagators of kappa and $K_0^*(1430)$ are affected by the effects of the $\pi K$ background. The background-resonance interference approximation extends the original concept of Ref.~\cite{GLSM_piK} to include complex coefficients, whereby the unitarized $I=1/2$, $J=0$ scattering amplitude can be approximated by:
\begin{equation}
T_0^{1/2}(s) = \frac{\rho(s)}{2} 
\left[
C_\kappa +
\sum_{i=1}^2\,
 {
	 \frac{A_{\kappa_i}\, \left( 2 m_{\kappa_i}\, \Gamma_{\kappa_i}\right)}
  { \rho_{0i} \left(m_{\kappa_i}^2 - s\right)  - i\, m_{\kappa_i} \Gamma_{\kappa_i} \rho_{0i}}
 }
\right]
\label{E_T012_rBW}
\end{equation}
where 
\begin{equation}
\rho(s) = \frac{q} 
{8 \pi \sqrt{s}}
= 
\frac{\sqrt{
	[s-(m_\pi+m_K)^2]
	[s-(m_\pi-m_K)^2]
}} 
{16 \pi s}
\label{E_rho_piK}
\end{equation} 
with $q$ the center of mass momentum, $\rho_{0i} = \rho(m_{\kappa_i}^2)$,  with $i=1, 2$ representing the light and the heavy kappas, and the complex constants $C_\kappa = C_{\kappa R} + i\, C_{\kappa I}$, $A_{\kappa_i} = A_{\kappa_i R} + i\, A_{\kappa_iI}$
are determined from fits to experimental data on $\pi K$ scattering.  The resonances have a modified Breit-Wigner form, and the two complex constants $A_{\kappa_1}$ and $A_{\kappa_2}$ give a measure of the deviation of each resonance from a Breit-Wigner shape.  In addition, these constants also contain information about the interference of the two resonances with each other, as well as the interference of each resonance with the background, thereby motivating our background-resonance interference approximation terminology. 
Moreover, since the fit parameters are constant complex numbers, both the real and the imaginary parts of each resonance contribute to the imaginary and real parts of the scattering amplitude:

\begin{gather}
{\rm Re}\left(T_0^{1/2}(s) \right)  = 
\frac{\rho(s)}{ 2} 
\left[
C_{\kappa R} + 
\sum_{i=1}^2\,
{
	\frac{
		2 m_{\kappa_i}\Gamma_{\kappa_i} \, \rho_{0i}\,
		\left[
			A_{\kappa_iR}\,  \left( m_{\kappa_i}^2 - s \right)
     	    - A_{\kappa_iI} \, m_{\kappa_i} \Gamma_{\kappa_i} \, 
     	 \right]
    }
{\rho_{0i}^2 \,\left( m_{\kappa_i}^2 - s \right)^2 + m_{\kappa_i}^2 \Gamma_{\kappa_i}^2\, \rho_{0i}^2}
}
\right]
\label{E_ReT012_rBW}
\\
{\rm Im}\left(T_0^{1/2} (s) \right)  = 
\frac{\rho(s)}{ 2} 
\left[
C_{\kappa I} + 
\sum_{i=1}^2\,
{
	\frac{
		2 m_{\kappa_i}\Gamma_{\kappa_i} \, \rho_{0i}\,
		\left[
		A_{\kappa_iI}\,  \left( m_{\kappa_i}^2 - s \right)
		+ A_{\kappa_iR} \, m_{\kappa_i} \Gamma_{\kappa_i} \, 
		\right]
	} 
	{\rho_{0i}^2 \,\left( m_{\kappa_i}^2 - s \right)^2 + m_{\kappa_i}^2 \Gamma_{\kappa_i}^2\, \rho_{0i}^2}
}
\right]
\label{E_ImT012_rBW}
\end{gather}

\begin{figure}[htb]
	\centering
	\includegraphics[width=0.32\columnwidth]{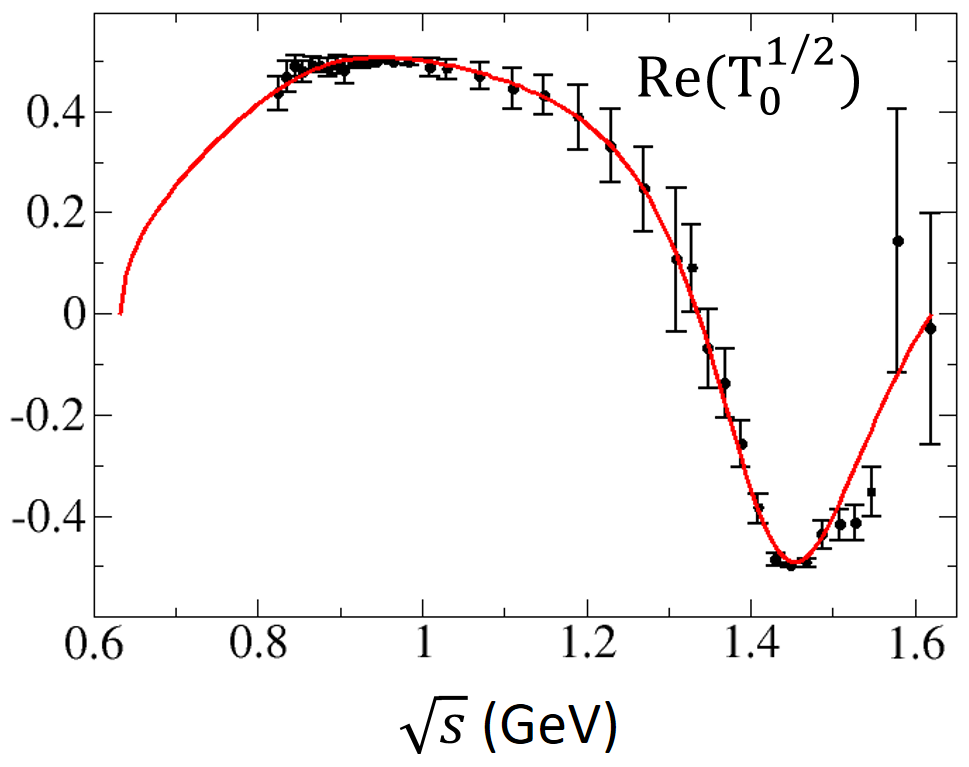}\hspace{0.02\columnwidth}
    \includegraphics[width=0.32\columnwidth]{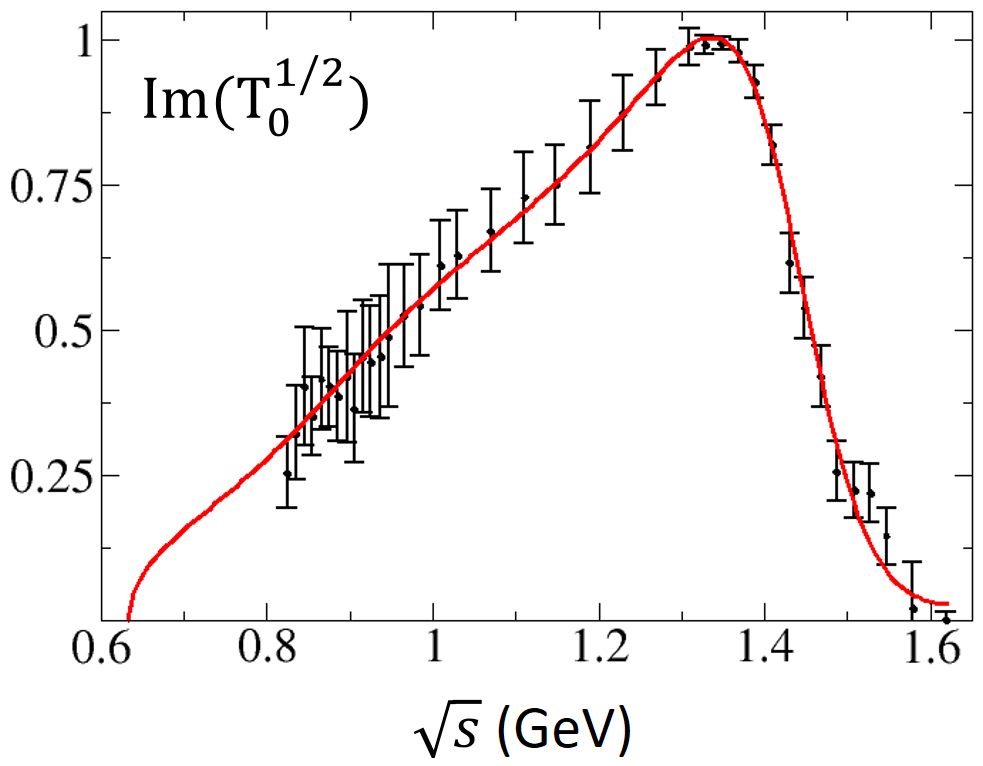}\hspace{0.02\columnwidth}
\vskip 1cm
    \includegraphics[width=0.32\columnwidth]{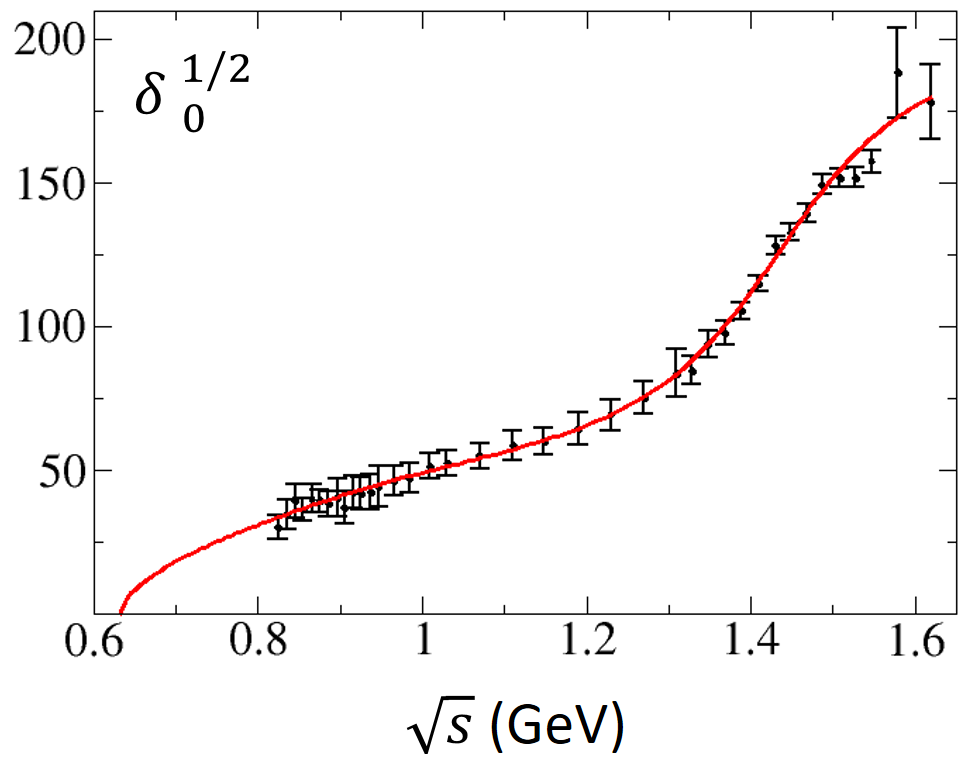}\hspace{0.02\columnwidth}
\caption{Fit of the background-resonance interference approximation model (\ref{E_T012_rBW})   
(solid red lines),  
to the experimental data \cite{Aston} (solid dots and error bars).
The first two figures show the real and imaginary parts of the $I=1/2$, $J=0$, $\pi K$  scattering amplitude, followed by the third figure that shows the phase shift.  }
 \label{F_rBW}
\end{figure}

Although this form (with only real parameters) was suggested in \cite{GLSM_piK} to only provide a reasonable approximation of the scattering amplitude, the extension to complex coefficients surprisingly results in nearly perfect fits to the experimental data on $\pi K$ scattering.   We perform a combined (unweighted) $\chi^2$ fit of the real and imaginary parts of  the scattering amplitude (\ref{E_ReT012_rBW}) to the experimental data \cite{Aston}: 
\begin{equation}
\chi^2 = 
\sum_{k=1}^{N}
\left[ 
{\rm Re} 
\left(
T_0^{1/2} (\sqrt{s_k})
\right) - 
{\rm Re} \left(
T_0^{1/2, {\rm Exp} } (\sqrt{s_k})
\right)
\right]^2
+ 
\left[ 
{\rm Im} \left(T_0^{1/2}(\sqrt{s_k})\right) - {\rm Im} \left(T_0^{1/2, {\rm Exp}}(\sqrt{s_k})\right)
\right]^2
\end{equation}
where $N$ is the number of data points. Since the real and the imaginary parts of $T_0^{1/2}$ depend linearly on the complex unknown constants $A_{\kappa_i}$ and $C_{\kappa_i}$, we can analytically determine these unknowns from 
\begin{equation}
\frac{\partial \chi^2}{\partial p_j} = 0 
\end{equation}
where $p_j = \{A_{\kappa_1R}, A_{\kappa_1I}, A_{\kappa_2 R}, A_{\kappa_2 I}, C_{\kappa R}, C_{\kappa I} \}$.  This results in a system of six linear equations in the six unknowns $p_j$:
\begin{equation}
Q_{ij}\,  p_j = V_i 
\end{equation}
where the matrix of coefficients $Q$ is:
\begin{equation}
Q_{ij} =  (-1)^{(i + j)} \left(
R_i^*  R_j  + R_i  R_j^* \right)
\end{equation}
with
\begin{equation}
R_1 = 
\frac{1}{2}
\, \sum_{k=1}^N\, \rho(s_k)
\left[
      {\rm Re} \left(T_{0 \kappa_1}^{1/2}(s_k)\right) + 
      i 
      \, {\rm Im} \left(T_{0 \kappa_1}^{1/2}(s_k)\right) 
\right]
\end{equation}
and $R_2 = -i\, R_1$, $R_3$ and $R_4$ are respectively obtained from $R_1$ and $R_2$ with the substitution $\kappa_1  \leftrightarrow \kappa_2$,  $R_5 = 1$ and $R_6 = -i$. Similarly, we find 
\begin{gather}
V_1 =
\sum_{k=1}^N\, \rho(s_k)\, \left\{
{\rm Im} \left(T_{0 }^{1/2,\, {\rm Exp}}(s_k)\right)  
{\rm Im} \left(T_{0 \kappa_1}^{1/2}(s_k)\right)
+
{\rm Re} \left(T_{0 }^{1/2,\, {\rm Exp}}(s_k)\right)  
{\rm Re} \left(T_{0 \kappa_1}^{1/2}(s_k)\right)
\right\} 
\\
V_2 =
\sum_{k=1}^N\, \rho(s_k)\, \left\{
{\rm Im} \left(T_{0 }^{1/2,\, {\rm Exp}}(s_k)\right)  
{\rm Re} \left(T_{0 \kappa_1}^{1/2}(s_k)\right)
-
{\rm Re} \left(T_{0 }^{1/2,\, {\rm Exp}}(s_k)\right)  
{\rm Im} \left(T_{0 \kappa_1}^{1/2}(s_k)\right)
\right\} \,,
\end{gather}
and $V_3$ and $V_4$ are respectively obtained from 
$ V_1$ and $V_2$ in which the substitution $\kappa_1 \leftrightarrow \kappa_2$ is made, and 
\begin{gather}
V_5 = \sum_{k=1}^N\, \rho(s_k)\, 
 {\rm Re} \left(T_{0 }^{1/2,\, {\rm Exp}}(s_k)\right)
 \\
V_6 = \sum_{k=1}^N\, \rho(s_k)\, 
 {\rm Im} \left(T_{0 }^{1/2,\, {\rm Exp}}(s_k)\right)~.
\end{gather}
The result of this fit is given in Table \ref{T_BW_EBW_fits}, column 1, and the resulting plots of the real and imaginary parts of the scattering amplitude as well as the phase shift are compared with the experimental data in Fig.~\ref{F_rBW} showing a remarkable agreement. 

\begin{figure}[ht]
	\centering
	\includegraphics[width=0.32\columnwidth]{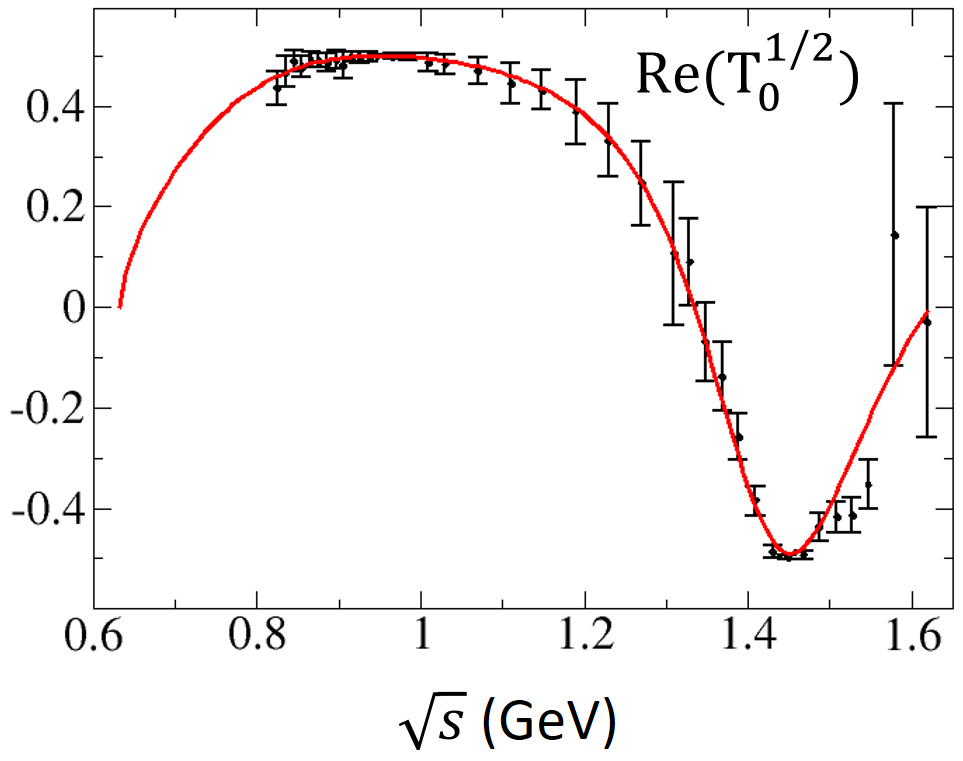}\hspace{0.02\columnwidth}
	\includegraphics[width=0.32\columnwidth]{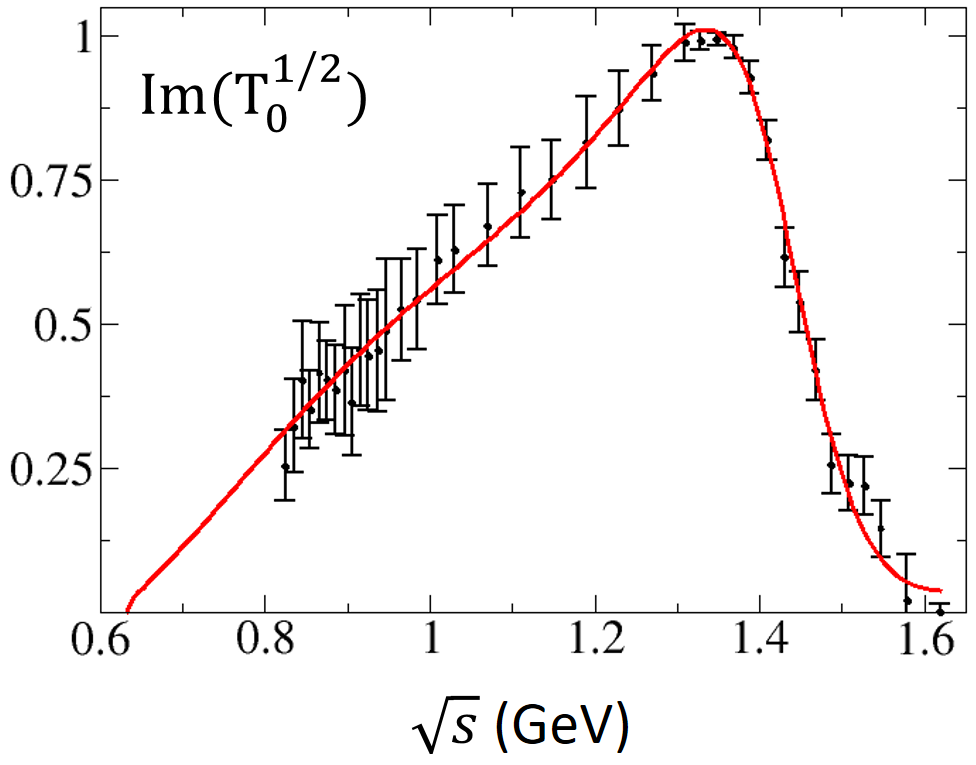}\hspace{0.02\columnwidth}
\vskip 1cm
 \includegraphics[width=0.32\columnwidth]{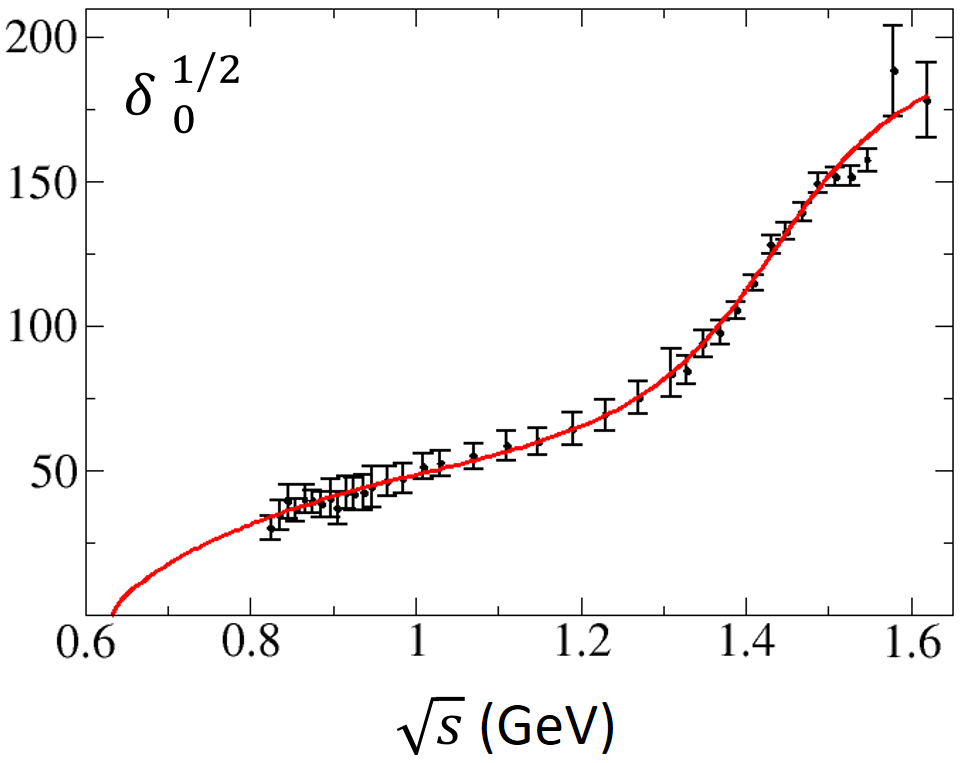}\hspace{0.02\columnwidth}
\caption{
Fit of the background-resonance interference approximation unitarized model (\ref{E_T012_rEBW})   
(solid red lines),  
to the experimental data \cite{Aston} (solid dots and error bars).
The first two figures show the real and imaginary parts of the $I=1/2$, $J=0$, $\pi K$  scattering amplitude, followed by the third figure that shows the phase shift. 
 }
	\label{F_rEBW}
\end{figure}

Our second parametrization of the scattering amplitude is similar to the parametrization discussed above and includes a constant complex background, but in this parametrization  each complex-weighted resonance contribution to the scattering amplitude is modified to be unitary.
In $\pi K$ scattering within the generalized linear sigma model \cite{GLSM_piK},  the contribution of kappa mesons in the $s$-channel (top-right diagram in Fig.~\ref{F_Feyn_diag}) to the isospin 1/2, spin 0 ``bare'' scattering amplitude is:
\begin{equation}
T_{0\, \kappa_i}^{
\frac{1}{2}
} = 
\frac{3 \rho(s)} 
{2} 
\, 
	\frac{\gamma_{\kappa_i \pi K}^2}
	{m_{\kappa_i}^2 - s}
\label{E_T012B}
\end{equation}
The ``bare'' amplitude should be unitarized in some fashion.   A generalized unitarization of (\ref{E_T012B}) around each pole is to add an energy-dependent imaginary part to the denominator \cite{00_BFMNS,00_SSESXF,94_AS}
\begin{equation}
T_{0 \, \kappa_i}^{\frac{1}{2}
} = 
\frac{3 \rho(s)} 
{2}
	\frac{\gamma_{\kappa_i \pi K}^2}
	{m_{\kappa_i}^2 - s - i m_{\kappa_i} G_i(s)}
\label{E_T012_unitary}
\end{equation} 
that satisfies the condition 
\begin{equation}
G_i(s) = 
\frac{3 \rho(s)} 
{2 m_{\kappa_i}}
\gamma_{\kappa_i \pi K}^2
\label{E_Gs}
\end{equation}
where $\gamma_{\kappa_i \pi K}$ is the coupling constant of $\kappa_i$ with $i=1,2$ to $\pi K$ computed in \cite{GLSM_piK}.  Since $G_i(m_{\kappa_i}^2) = \Gamma_{\kappa_i} = \Gamma [\kappa_i \rightarrow \pi K]$, 
we can also rewrite (\ref{E_T012_unitary}) in the form 
\begin{equation}
T_{0 \, \kappa_i}^{\frac{1}{2}
} =
\frac{\rho(s)\,m_{\kappa_i}  \Gamma_{\kappa_i}  }
	{\rho_{0i}\, \left( m_{\kappa_i}^2 - s \right) - i\, m_{\kappa_i} \Gamma_{\kappa_i} \, \rho(s)}
\label{E_T012_kappai}
\end{equation}
Using this energy-dependent unitarized  form for the contribution of each kappa to the scattering amplitude, we can then use the background-resonance interference approximation to calculate,  similar to  Eq.~(\ref{E_T012_rBW}),  the total scattering amplitude as the sum of a constant complex background plus the two $s$-channel poles of kappas:   
\begin{equation}
T_0^{1/2} (s) = \frac{\rho (s)} 
{2} 
\left[
C_\kappa +
\sum_{i=1}^2\,
\frac{
A_{\kappa_i}\, 	\left(2 m_{\kappa_i}  \Gamma_{\kappa_i} \right)}
	{\rho_{0i} \, \left( m_{\kappa_i}^2 - s \right) - i\, m_{\kappa_i} \Gamma_{\kappa_i} \, \rho(s)}
\right]\,.
\label{E_T012_rEBW}
\end{equation}
The real and the imaginary parts of the scattering amplitude are then fitted to experimental data \cite{Aston} and the result is given in column~2 of Table \ref{T_BW_EBW_fits} and the resulting graphs are displayed in Fig.~\ref{F_rEBW}.  We see in Table~\ref{T_BW_EBW_fits} that the  fitted parameters are relatively stable in these two models.

\begin{table}[htb]
\renewcommand{\arraystretch}{1.5}
\centering
	\begin{tabular}{c|c|c}
		\hline
		\rule{0pt}{3ex}   
	Parameter & Model (\ref{E_T012_rBW})   &   Model (\ref{E_T012_rEBW})   
		\\[2pt]
		\hline
		$A_{\kappa_1}$	 &  $-0.0560 - i\, 0.144$    &  $-0.0559 - i\,  0.150$ 
		\\
		$A_{\kappa_2}$  &  $0.385 + i\,  0.899$ &  $0.382 + i\, 0.900$ 
		\\
		$C_\kappa$  &  $47.853 + i\,  33.116$ &  $46.861 + i\, 33.386 $
		\\
			\hline
	$\Delta \delta_0^1$ &  0.026            &    0.028
	\end{tabular}
	\caption{The  complex parameters of the background-resonance interference approximation models (\ref{E_T012_rBW})
    and (\ref{E_T012_rEBW}) determined from fits to experimental data \cite{Aston}.  The goodness of fit $\Delta \delta_0^1$ [defined in (\ref{E_pik_fit_goodness})] is also provided for each model.
	}
	\label{T_BW_EBW_fits}
\end{table}

We measure the goodness of the fits by calculating the average relative error between model computation of phase shift and the central values of the experimental data on phase shift:
\begin{equation}
\Delta \delta_0^1 = 
\frac{1}{N}\, 
\sum_k^N 
{ 
	\frac{\left| \delta_0^{1, {\rm Theo.} } (s_k)- \delta_0^{1, {\rm Exp.}} (s_k)  \right|}{\delta_0^{1, {\rm Exp.}} (s_k)}
}
\label{E_pik_fit_goodness}
\end{equation}
These errors are computed and given in the last row of Table \ref{T_BW_EBW_fits}.  These are much smaller than the average relative experimental uncertainty given by
\begin{equation}
\Delta \delta_0^{1, Exp.} = 
\frac{1}{N}
\, 
\sum_k^N 
{ 
	\frac{\Delta \delta_0^{1, {\rm Exp.}} (s_k)}{\delta_0^{1, {\rm Exp.}} (s_k)}
} = 0.09 
\end{equation}

\subsection{The $a_0$ system}
\label{a0_subsection}

\begin{figure}[ht]
	\centering
	\includegraphics[width=0.4\columnwidth]{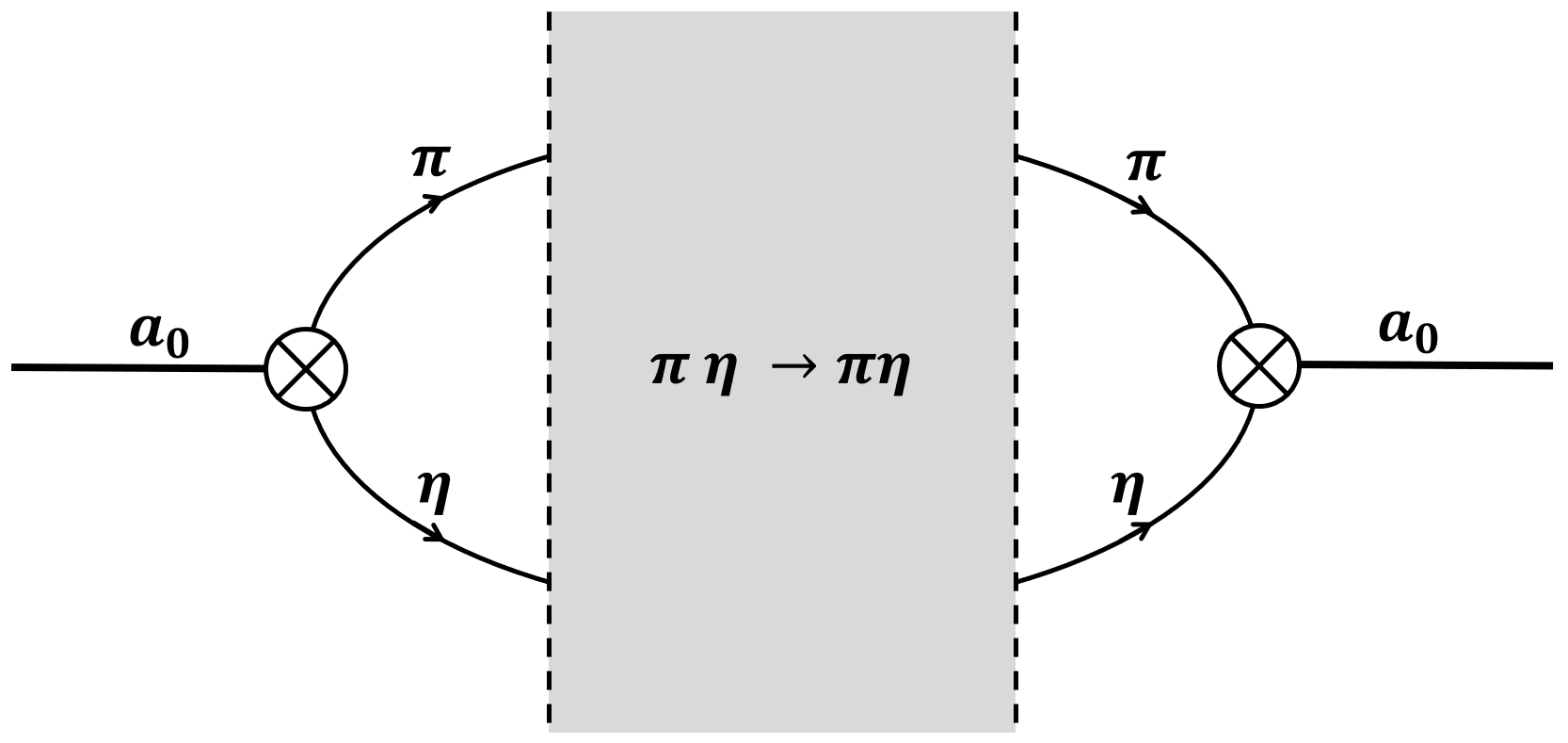}\hspace{0.02\columnwidth}
	\caption{Inclusion of $\pi \eta$ intermediate states in the $a_0$ correlator.
	}   
	\label{F_a0_correlator}
\end{figure}

\begin{figure}[ht]
	\centering
	\includegraphics[width=0.4\columnwidth]{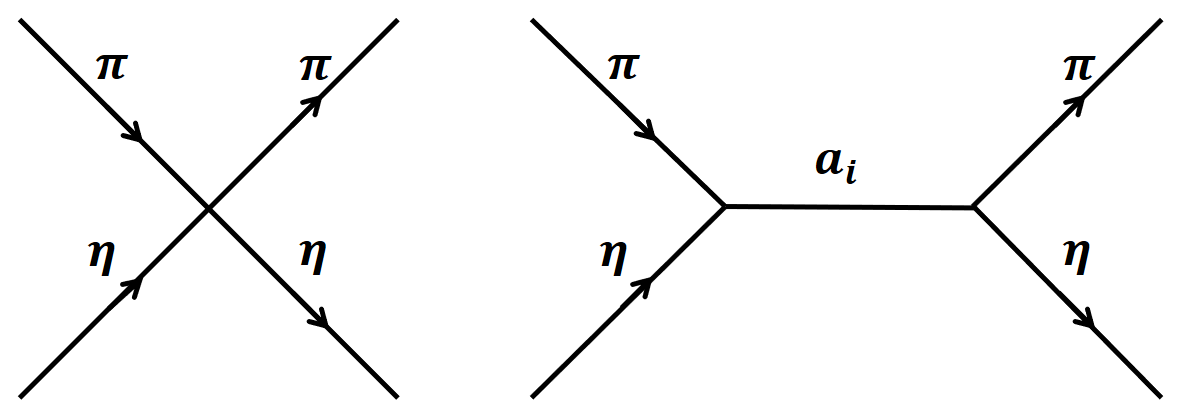}\hspace{0.02\columnwidth}
	\vskip 1cm
	\includegraphics[width=0.4\columnwidth]{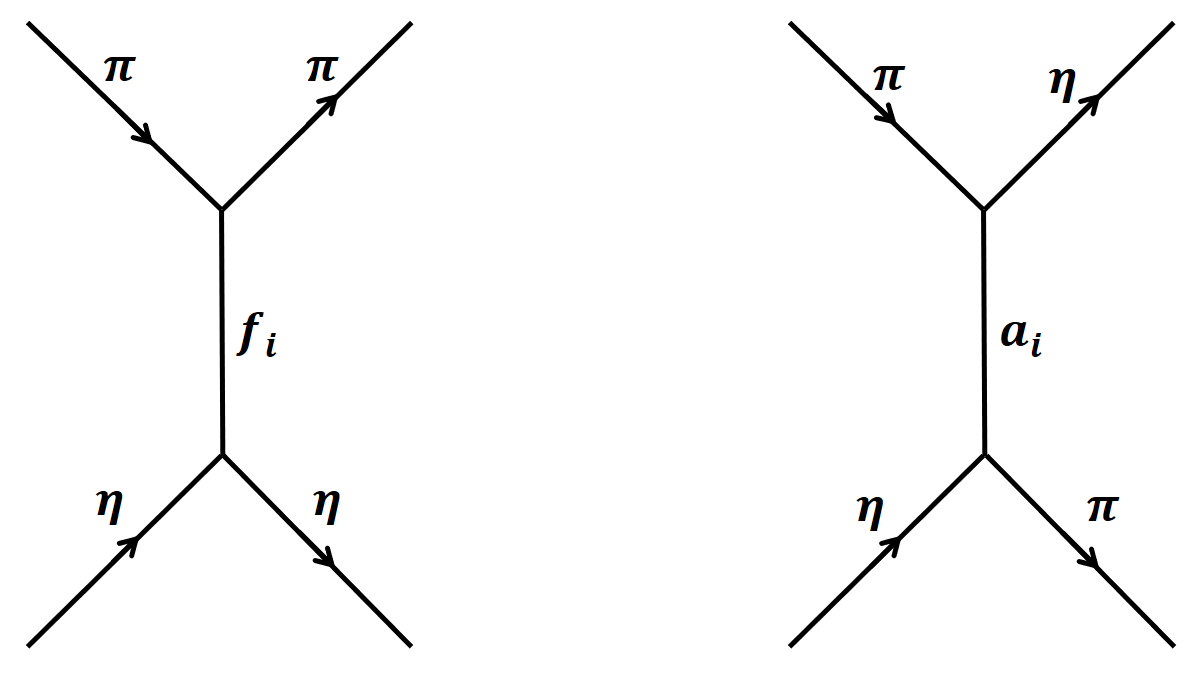}\hspace{0.02\columnwidth}
	\caption{Tree level Feynman diagrams for $\pi \eta$ scattering amplitude in the generalized linear sigma model.
	}
	\label{F_Feyn_diag_pieta}
\end{figure}

Similar to the case just discussed for modeling the kappa shape, we model the $a_0$ shape by inclusion of $\pi\eta$ intermediate states in the hadronic correlator as shown in Fig.~\ref{F_a0_correlator} (in principle, the $K {\bar K}$ and the $\pi \eta'$ channels also contribute, but in the first approximation we only include the $\pi\eta$ channel in this analysis).
The Feynman diagrams contributing to the tree-level $\pi \eta$ scattering amplitude are given in Fig.~\ref{F_Feyn_diag_pieta} and include a four-point contribution together with contributions of the isotriplet and isosinglet 
scalars as shown.

\begin{figure}[htb]
	\centering
	\includegraphics[width=0.32\columnwidth]{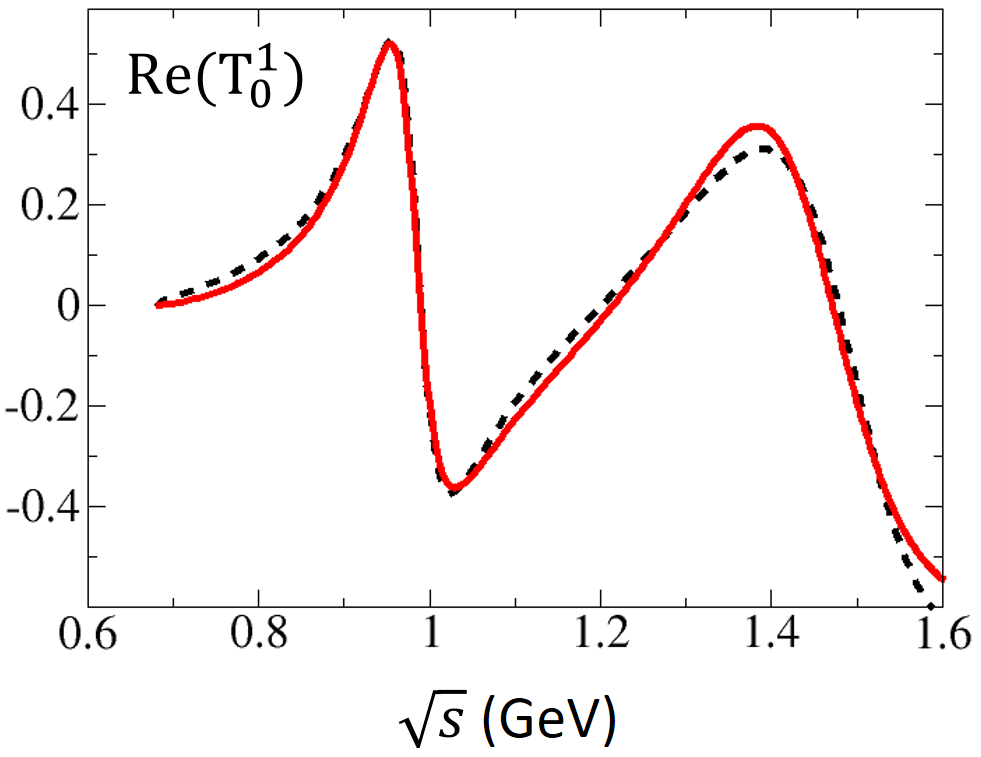}\hspace{0.02\columnwidth}
    \includegraphics[width=0.32\columnwidth]{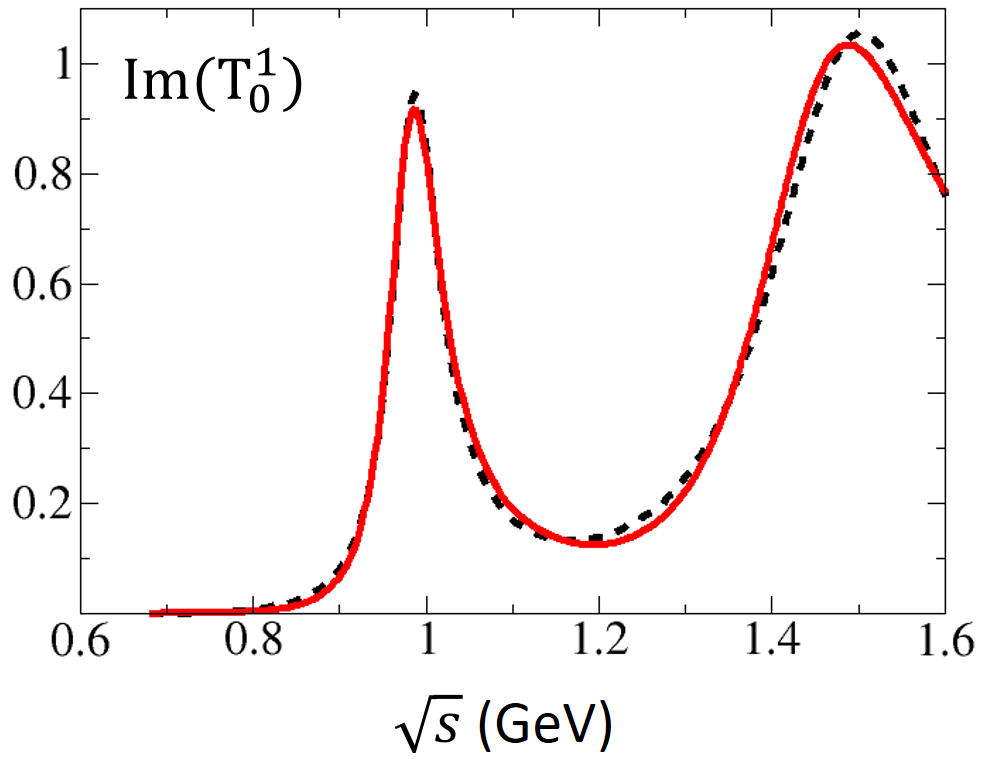}\hspace{0.02\columnwidth}
    \vskip 1cm
    \includegraphics[width=0.32\columnwidth]
{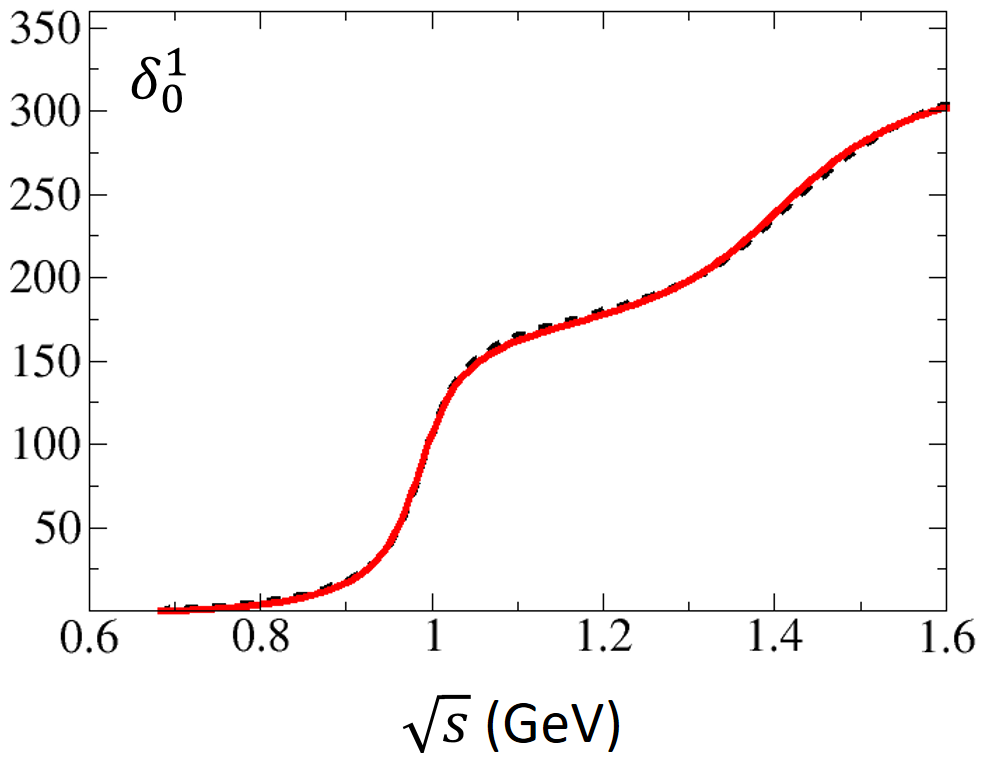}\hspace{0.02\columnwidth}
\caption{
Fit of the background-resonance interference approximation model (\ref{E_T01_rBW}) (solid red lines),  
to the theoretical predictions of \cite{00_BFS61} (dashed  lines).
The first two figures show the real and imaginary parts of the $I=1$, $J=0$, $\pi \eta$  scattering amplitude, followed by the third figure that shows the phase shift.  
	}
	\label{F_pieta_fit_BW}
\end{figure}

The unitarized  $I=1$, $J=0$ scattering amplitude can be parameterized in the background-resonance interference approximation as: 
\begin{equation}
T_0^1(s) = \frac{\rho(s)}{2} 
\left[
C_a +
\sum_{i=1}^2\,
 {
	 \frac{A_{a_i}\, \left( 2 m_{a_i}\, \Gamma_{a_i}\right)}
  { \rho_{0i} \left(m_{a_i}^2 - s\right)  - i\, m_{a_i} \Gamma_{a_i} \rho_{0i}}
 }
\right]
\label{E_T01_rBW}
\end{equation}
where 
\begin{equation}
\rho(s) = \frac{q}
{8 \pi \sqrt{s}}
= 
\frac{\sqrt{
	\left[s - (m_\pi + m_\eta)^2\right]
	\left[s-  (m_\pi - m_\eta)^2\right] 
}}
{16 \pi s}
\label{E_rho_pieta}
\end{equation} 
with $q$ the center of mass momentum, 
 $\rho_{0i} = \rho(m_{a_i}^2)$,  with $i=1, 2$ representing the light and the heavy $a_0$, and the a priori unknown  complex constants $C_a = C_{aR} + i\, C_{aI}$, $A_{a_i} = A_{a_i R} + i\, A_{a_i I}$ 
 have to be determined from fits to the scattering amplitude.    Since there are no experimental data on $\pi\eta$ scattering,  the constants $C_a$ and $A_{a_i}$ have to be determined by fitting (\ref{E_T01_rBW}) to a theoretical prediction for this scattering amplitude. A comparison between different theoretical predictions is made in \cite{GLSM_pieta}  showing some discrepancy among different approaches.  We fit to the work of \cite{00_BFS61} which qualitatively shows an average behavior among different predictions. Using the same fitting method described for the kappa system, the result of our background-resonance interference approximation fit to the real and imaginary parts of the $J=0$, $I=1$ $\pi\eta$ scattering amplitude of Ref.~\cite{00_BFS61} is given in Table \ref{T_BW_EBW_a0_fits}, and the plots of these fits together with the phase shift are given in Fig.~\ref{F_pieta_fit_BW}.

\begin{figure}[htb]
	\centering
	\includegraphics[width=0.32\columnwidth]{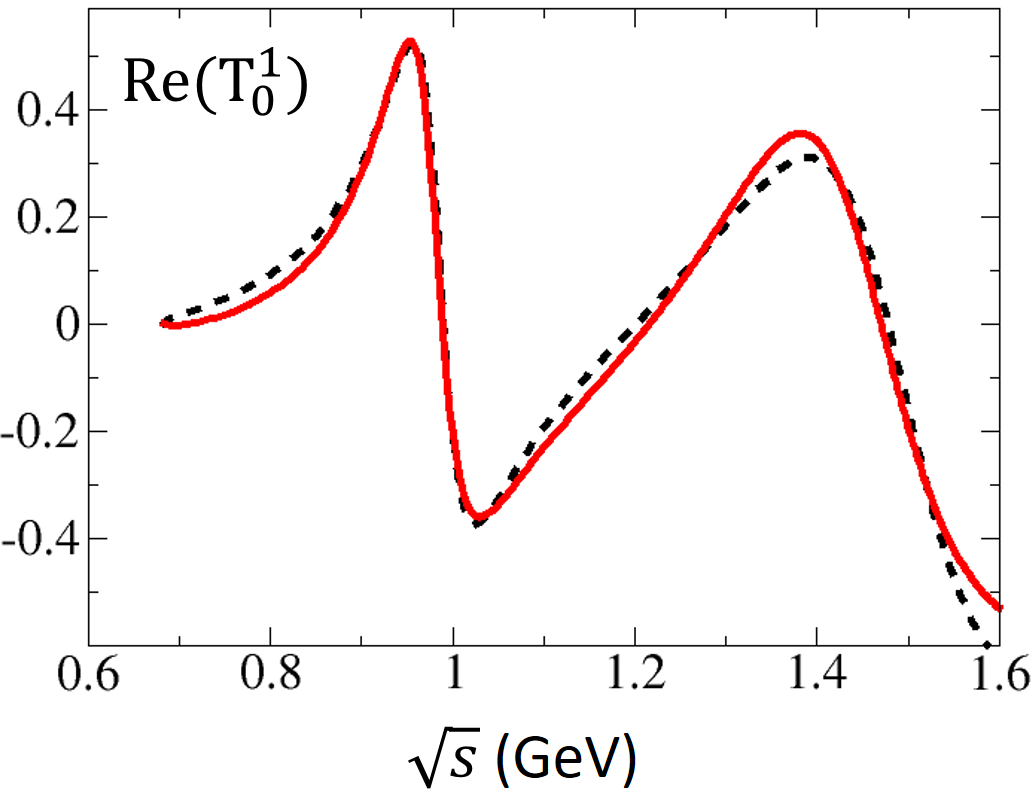}\hspace{0.02\columnwidth}
    \includegraphics[width=0.32\columnwidth]{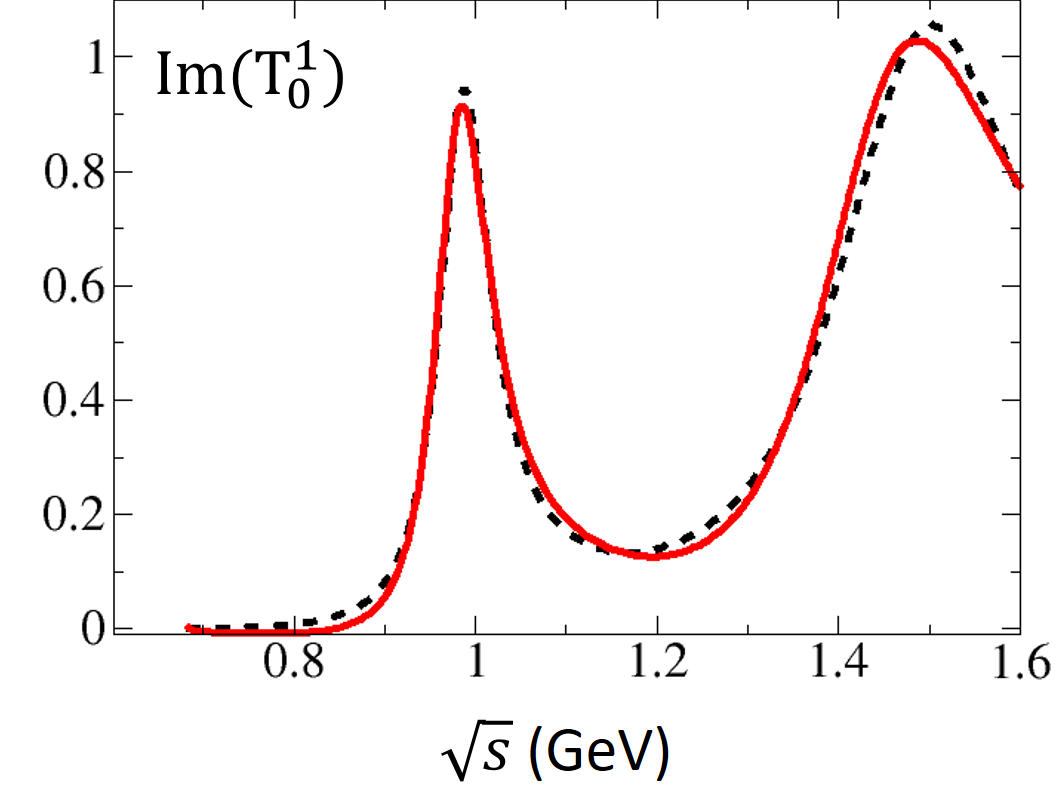}\hspace{0.02\columnwidth}
\vskip 1cm
    \includegraphics[width=0.32\columnwidth]{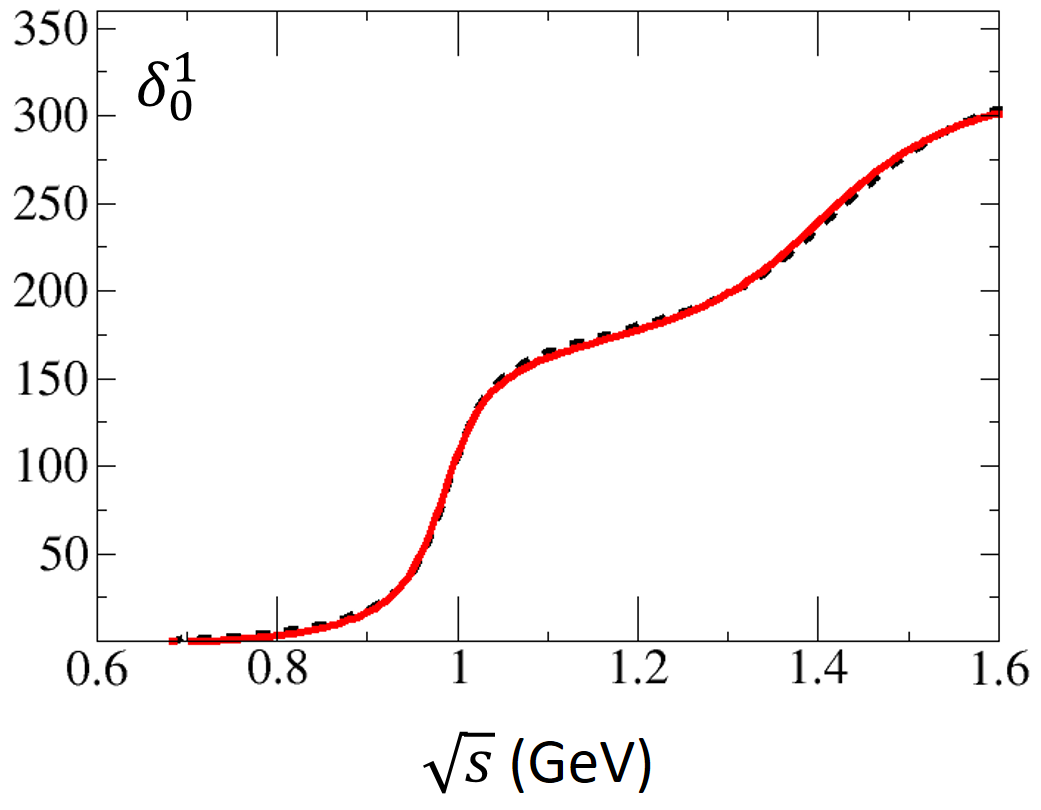}\hspace{0.02\columnwidth}
\caption{Fit of the background-resonance interference approximation model (\ref{E_T01_a0_EBW})   
(solid red lines),  
to the theoretical predictions of \cite{00_BFS61} (dashed  lines).
The first two figures show the real and imaginary parts of the $I=1$, $J=0$, $\pi \eta$  scattering amplitude, followed by the third figure that shows the phase shift. }
	\label{F_pieta_fit_EBW}
\end{figure}

 We also examine the background-resonance interference approximation fit with local unitary modeling of the amplitude 
\begin{equation}
T_0^1(s) = \frac{\rho(s)}{2} 
\left[
C_a +
\sum_{i=1}^2\,
	 \frac{A_{a_i}\, \left( 2 m_{a_i}\, \Gamma_{a_i}\right)}
  { \rho_{0i} \left(m_{a_i}^2 - s\right)  - i\, m_{a_i} \Gamma_{a_i} \rho(s)}
\right]
\label{E_T01_a0_EBW}
\end{equation}
 the result of which is also displayed in Table \ref{T_BW_EBW_a0_fits} showing a negligible difference between the two fits. The plots of the real and imaginary parts of $\pi\eta$  scattering amplitude as well as the phase shift obtained in this fit are compared with those of Ref.~\cite{00_BFS61} in Fig.~\ref{F_pieta_fit_EBW}.
 
\begin{table}[ht]
\renewcommand{\arraystretch}{1.5}
\centering
	\begin{tabular}{c|c|c}
		\hline
		\rule{0pt}{3ex}   
	Parameter & Model (\ref{E_T01_rBW}) &   Model (\ref{E_T01_a0_EBW})   
		\\[2pt]
		\hline
		$A_{a_1}$	 &  $0.864 - i\, 0.298$    &  $0.872 - i\,  0.288$ 
		\\
		$A_{a_2}$  &  $0.726 - i\,  0.641$ &  $0.710 - i\, 0.635$ 
		\\
		$C_a$  &  $-42.7 + i\,  17.6$ &  $-43.0 + i\, 19.9$
	\end{tabular}
	\caption{The  complex parameters of the background-resonance interference approximation models (\ref{E_T01_rBW}) and (\ref{E_T01_a0_EBW}) determined from fits to theoretical prediction of \cite{00_BFS61}.}
	\label{T_BW_EBW_a0_fits}
\end{table}

\section{Scale Factor Methodology Review}
\label{scale_factor_section}

In this section we review the scale factor methodology that relates chiral-Lagrangian fields to QCD composite operators (currents) \cite{CLQCDSR_2016,CLQCDSR:2019_Proc,CLQCDSR_2020,Fariborz:2019zht}.  
The bare (un-mixed) mesonic chiral fields $M$ and $M'$ are represented as
\begin{equation}
M  =  S + i\phi \,,~M'  =  S' + i \phi' 
\end{equation}
where  $S$ and  $\phi$ are respectively bare quark-antiquark scalar nonet and bare quark-antiquark pseudoscalar nonet.  Similarly,  $S'$ and $\phi'$ are unmixed scalar and pseudoscalar four-quark nonets made of two quarks and two  antiquarks.
\begin{gather}
\!S=
\begin{pmatrix}
S_1^1 & a_0^+ & \kappa^+  \\
a_0^- & S_2^2 & \kappa^0 \\
\kappa^- & {\bar \kappa}^0 & S_3^3
\end{pmatrix}, 
~
S' =
\begin{pmatrix}
{S'}_1^1 & {a'}_0^+ & {\kappa'}^+  \\
{a'}_0^- & {S'}_2^2 & {\kappa'}^0 \\
{\kappa'}^- & {\bar {\kappa'}}^0 & {S'}_3^3 \\
\end{pmatrix}
\label{SpMES_new}
\end{gather}
and similarly for the pseudoscalar components in $\phi$ and $\phi'$. The chiral nonets $M$ and $M'$ have the same chiral-transformation properties but behave differently under $U_A(1)$ transformations
\begin{gather}
M  \rightarrow U_L \,  M \,  U_R^\dagger\,,  \qquad M\rightarrow e^{2i\nu}M\nonumber \\
M'  \rightarrow U_L \,  M' \,  U_R^\dagger\,, ~\quad  M'\rightarrow e^{-4i\nu}M'~,
\label{M_trans_new}
\end{gather}
which allows us to classify two-quark and four-quark operators sharing the transformation properties \eqref{M_trans_new}.

We can also define the chiral nonets at the quark-level.  For example, the quark-antiquark operators can be represented as 
\begin{equation}
(\m)_a^b = ({\bar q}_R)^b ({q_L})_a  \Rightarrow \left(S_{\rm QCD}\right)_a^b =q_a(x) {\bar q}^b(x)
\end{equation}
where $a$ and $b$ are flavor indices. In a similar way 
$M'_{\,\rm QCD}$ is composed of four-quark operators, but given the various
possibilities available the specific forms will be specified below as needed.   Thus the chiral nonets $\m$ and $M'_{\,\rm QCD}$ are populated by QCD currents (composite operators)
\begin{gather}
\m  =  S_{\rm QCD} + i\phi_{\rm QCD} \,,~M_{\rm QCD}'  =  S'_{\rm QCD} + i \phi'_{\rm QCD} 
\label{M_QCD_J}\\
\!\!S_{\rm QCD}=
\begin{pmatrix}
J_1^{11} & J_1^{a_0^+} & J_1^{\kappa^+}  \\
J_1^{a_0^-} & J_1^{2^2} & J_1^{\kappa^0} \\
J_1^{\kappa^-} & J_1^{{\bar \kappa}^0} & J_1^{33}
\end{pmatrix}, 
~
S'_{\rm QCD} =
\begin{pmatrix}
J_2^{11} & J_2^{{a'}_0^+} & J_2^{{\kappa'}^+}  \\
J_2^{{a'}_0^-} & J_2^{22} & J_2^{{\kappa'}^0} \\
J_2^{{\kappa'}^-} & J_2^{{\bar {\kappa'}}^0} & J_2^{33} \\
\end{pmatrix}
\label{SpMQCD_new}
\end{gather}
where the generic notation $J_1$  represents a two-quark current and $J_2$ represents a four-quark current.

We now connect the appropriate QCD operators to their corresponding mesonic fields by introducing scale factor matrices  $I_{M}$ and $I_{M'}$ via~\cite{CLQCDSR_2016,CLQCDSR:2019_Proc,CLQCDSR_2020,Fariborz:2019zht}
\begin{equation}
M=I_{M} M_{\rm QCD}\,,~M'=I_{M'} M'_{\,\rm QCD}~.
\label{M_scale_new}
\end{equation}
As outlined in Refs.~\cite{CLQCDSR_2016,CLQCDSR:2019_Proc,CLQCDSR_2020,Fariborz:2019zht}, the scale factor matrices are constrained by the chiral symmetry transformations \eqref{M_trans_new} 
\begin{gather}
 [U_R,I_{M}]= [U_L,I_{M}]=0~,
  \label{I_M_new}\\
 [U_R,I_{M'}]= [U_L,I_{M'}]=0~,
 \label{I_Mp_new}
\end{gather}
implying that the scale factor matrices have a simple form~\cite{CLQCDSR_2016,CLQCDSR:2019_Proc,CLQCDSR_2020,Fariborz:2019zht}
\begin{equation}
I_M=-\frac{m_q}{\Lambda^3}\times \mathds{1}\,,~I_{M'}=\frac{1}{{\Lambda'}^5}\times  \mathds{1}\, ,
\label{scale_factors_new}
\end{equation}
where the energy-independent (constant) scale factors $\Lambda$ and $\Lambda'$ have dimensions of energy and the quark mass factor $m_q=(m_u+m_d)/2$ incorporates the necessary renormalization-group behaviour associated with the QCD operators. The result \eqref{scale_factors_new} exhibits the crucial property of universality; the scale factor parameters   $\Lambda$ and $\Lambda'$ must be identical for all members of the nonet.  It is quite remarkable that only two energy-independent parameters characterize the relationship between chiral Lagrangian fields and their QCD current counterparts. Analysis of the isotriplet $a_0$ and isodoublet $K_0^*$ sectors has confirmed this universality \cite{CLQCDSR_2020}, providing important evidence for the scale factor relationship (\ref{M_scale_new},\ref{scale_factors_new}) between chiral Lagrangian fields and QCD currents~\cite{CLQCDSR_2016,CLQCDSR:2019_Proc,CLQCDSR_2020,Fariborz:2019zht}. 

For the 
$K_0^*$ isodoublet system, the detailed implementation of the relation between the physical (mixed) mesonic fields and their unmixed QCD currents using Chiral Lagrangian rotation matrix $L_\kappa$ and the scale factor matrix $I_\kappa$ is given by~\cite{CLQCDSR_2016,CLQCDSR:2019_Proc,CLQCDSR_2020,Fariborz:2019zht}

\begin{gather}
{\bf K}=
\begin{pmatrix}
K_0^*(700)\\
K_0^*(1430)
\end{pmatrix}
= L_\kappa^{-1}
\begin{pmatrix}
S^3_2\\
\left(S'\right)^3_2
\end{pmatrix}
=
{L_\kappa^{-1} I_\kappa J_\kappa^{\rm QCD} }
\label{K_def_new}
\\[5pt]
L^{-1}_\kappa=\begin{pmatrix}
\cos\theta_\kappa & -\sin\theta_\kappa
\\
\sin\theta_\kappa & \cos\theta_\kappa
\end{pmatrix}
\,,~I_\kappa =
\begin{pmatrix}
\frac{-m_q}{\Lambda^3} &0 \\
0 & \frac{1}{{{\Lambda'}^5}}
\end{pmatrix}
~,
\label{I_matrices_new}
\\[5pt]
J_\kappa^{\rm QCD}=\begin{pmatrix}
J^{\kappa}_1\\
J^{\kappa}_2
\end{pmatrix}
\,,~
J^{\kappa}_1=\bar ds 
\label{J1_kappa}
\\[5pt]
J^{\kappa}_2=\sin(\phi) u^T_\alpha C\gamma_\mu\gamma_5 s_\beta\left(\bar d_\alpha\gamma^\mu\gamma_5 C\bar u_\beta^T-\alpha\leftrightarrow \beta \right)
+\cos(\phi) d^T_\alpha C\gamma_\mu s_\beta\left(\bar d_\alpha\gamma^\mu C\bar u_\beta^T+\alpha\leftrightarrow \beta \right)
\label{J2_kappa}
\end{gather}
where $C$ is the charge conjugation operator and $\cot\phi=1/\sqrt{2}$ \cite{Chen:2007xr,Chen:2006hyp}. Similarly, the analogous expression for the  $a_0(980)$ and $a_0(1450)$ isotriplet system is \cite{CLQCDSR_2016,CLQCDSR:2019_Proc,Fariborz:2019zht}

\begin{gather}
{\bf A}=
\begin{pmatrix}
a_0^0(980)\\
a_0^0(1450)
\end{pmatrix}
= L_a^{-1}
\begin{pmatrix}
\frac{S_1^1 - S_2^2}{\sqrt{2}}\\
\frac{{S'}_1^1 - {S'}_2^2}{\sqrt{2}}
\end{pmatrix}
=
L_a^{-1} I_a J_a^{\rm QCD} 
\label{A_def_new}
\\[5pt]
L^{-1}_a=\begin{pmatrix}
\cos\theta_a & -\sin\theta_a
\\
\sin\theta_a & \cos\theta_a
\end{pmatrix}
\,,~I_a =I_\kappa=
\begin{pmatrix}
\frac{-m_q}
{\Lambda^3} &0 \\
0 & \frac{1}
{{\Lambda'}^5}
\end{pmatrix}
\\[5pt]
J_a^{\rm QCD} 
=\begin{pmatrix}
J^a_1\\
J^a_2
\end{pmatrix}\,,~
J^a_1=\left(\bar u u-\bar d d\right)/\sqrt{2}
\label{J_a_QCD_new}
\\[5pt]
J^a_2=\frac{\sin\phi}{\sqrt{2}}d^T_\alpha C\gamma_\mu\gamma_5 s_\beta\left(\bar d_\alpha\gamma^\mu\gamma_5 C\bar s_\beta^T-\alpha\leftrightarrow \beta \right)
+\frac{\cos\phi}{\sqrt{2}}d^T_\alpha C\gamma_\mu s_\beta\left(\bar d_\alpha\gamma^\mu C\bar s_\beta^T+\alpha\leftrightarrow \beta \right)
- u\leftrightarrow d\,,
\label{J2_a}
\end{gather}
where $C$ is the charge conjugation operator and $\cot\phi=1/\sqrt{2}$ \cite{Chen:2007xr,Chen:2006hyp}.

The key element of the QCD sum-rule analysis is the recognition that Eqs.~\eqref{K_def_new} and \eqref{A_def_new} contain the physical hadronic fields ${\bf H}=\{ {\bf K}\,, {\bf A} \}$, and hence their hadronic correlation function is diagonal
\begin{gather}
\Pi^{\rm H}_{ij}  \left(Q^2 =-q^2\right)
=i\int d^4x\, e^{iq\cdot x}
\langle 0| {\rm T} \left[ {\bf H}_i (x) {\bf H}_j(0) \right] |0 \rangle=0\,, ~i\ne j\,.
\label{had_diag_property}
\end{gather} 
Thus the combination of QCD currents on the RHS of \eqref{K_def_new} and \eqref{A_def_new} define projected physical currents $J^P$ and an associated  physically-projected  QCD correlation function matrix
$\Pi^P(Q^2)$~\cite{CLQCDSR_2016,CLQCDSR:2019_Proc,CLQCDSR_2020,Fariborz:2019zht}
\begin{gather}
J_s^P = L_s^{-1} I_s J_s^{\rm QCD}
\,,~s=\{\kappa\,,a\} 
\\[5pt]
\Pi^P(Q^2) = {\widetilde {\cal T}}^s \Pi^{\rm QCD}(Q^2)  {\cal T}^s\,,~~{\cal T}^s= I_s \, L_s
 \label{phys_corr_new}
\\[5pt]
 \Pi^{\rm QCD}_{mn}\left(Q^2=-q^2\right)= i\int d^4x\, e^{iq\cdot x}\langle 0| {\rm T}  \left[ J^{\rm QCD}_m (x) J_n^{\rm QCD}(0)^\dagger \right] |0 \rangle
\end{gather}
where  ${\widetilde {\cal T}}$ denotes the transpose of the matrix ${\cal T}$.
Using quark-hadron duality to equate the QCD and hadronic correlation functions
\begin{equation}
\Pi^H(Q^2) =\Pi^P(Q^2)= {\widetilde {\cal T}}^s \Pi^{\rm QCD}(Q^2)  {\cal T}^s\,,
\label{correlator_sum_rule_new}
\end{equation}
implies that $\Pi^P$ inherits the diagonal property \eqref{had_diag_property} of the hadronic correlator, leading the the following relation between elements of the QCD correlation function
\begin{equation}
 \Pi_{12}^{\rm QCD} = -
\left[
{
 \frac{    {\widetilde {\cal T}}^s_{11} \Pi_{11}^{\rm QCD} {\cal T}^s_{12}
    + {\widetilde {\cal T}}^s_{12} \Pi_{22}^{\rm QCD} {\cal T}^s_{22}
 }
 {{\widetilde {\cal T}}^s_{11}  {\cal T}^s_{22} + {\widetilde {\cal T}}^s_{12}  {\cal T}^s_{12}
 }
}
\right]~.
\label{Pi_12_constraint}
\end{equation}
Because the off-diagonal QCD correlation function between a two-quark and four-quark current is unknown (and very difficult to compute because its leading contribution is four-loops), Eq.~\eqref{Pi_12_constraint} was used as a QCD input for the scale-factor analysis of 
Refs.~\cite{CLQCDSR_2016,CLQCDSR:2019_Proc,CLQCDSR_2020,Fariborz:2019zht}. In the next section, we identify some challenges associated with this approach and develop an alternative analysis methodology to address these challenges.

\section{Revised Scale Factor Methodology}
\label{new_scale_methodology_sec}

The off-diagonal constraint \eqref{Pi_12_constraint} is very easily implemented in the two-dimensional ($a_0,K_0^*)$ isospin sub-systems outlined in the previous section.  However, with an ultimate goal of a complete study of the scalar nonet including an additional gluonium component within the scale-factor framework, it is necessary to consider higher-dimensional isospin subsystems. Extending the diagonal physical constraint  \eqref{had_diag_property} to a three-dimensional subsystem already leads to an impractical number of constraints, a situation that becomes even worse for the five-dimensional system for the scalar isoscalars (with gluonium).    

With the goal of extending our scale-factor framework to higher-dimensional systems, we rearrange the mixing angle matrix in Eqs.~(\ref{K_def_new},\ref{A_def_new}) to obtain
\begin{equation}
L_s{\bf H}=I_s J_s^{\rm QCD}
\,,~s=\{\kappa\,,a\} \,. 
\label{current_unmixed}
\end{equation}
Because the scale factor matrix $I_s$ is diagonal,  the components of the QCD currents can be immediately determined as mixtures of physical chiral Lagrangian hadronic fields.  Diagonal QCD correlation functions can then be calculated using the constraint on the off-diagonal elements \eqref{had_diag_property} for the physical hadronic field correlator, resulting in the following relations 
\begin{gather}
 \frac{m_q^2}{\Lambda^6}\Pi^{\rm QCD}_{11}\left(Q^2\right) =\cos^2{\theta_s}\Pi^H_{11}\left(Q^2\right)+  \sin^2{\theta_s}\Pi^H_{22}\left(Q^2\right)\,,
 \label{Pi_QCD_11}
 \\
 \frac{1}{\left(\Lambda'\right)^{10}}\Pi^{\rm QCD}_{22}\left(Q^2\right) =\sin^2{\theta_s}\Pi^H_{11}\left(Q^2\right)+  \cos^2{\theta_s}\Pi^H_{22}\left(Q^2\right)\,.
 \label{Pi_QCD_22}
\end{gather}
Although we do not expect the relations Eqs.~\eqref{Pi_QCD_11} and \eqref{Pi_QCD_22} to be valid at all $Q^2$,  they can be used to obtain quark-hadron duality relations underlying the construction of QCD sum-rules.  Similarly, we expect the QCD off-diagonal constraint \eqref{Pi_12_constraint} to be valid in the sense of a duality relation. Because our revised scale factor methodology avoids use of the QCD constraint \eqref{Pi_12_constraint}, it thus eliminates a possible source of theoretical uncertainty.

With multiple hadronic states explicitly present in Eqs.~\eqref{Pi_QCD_11} and \eqref{Pi_QCD_22}, we use Gaussian QCD sum-rules 
\cite{gauss,harnett_quark}
because of their ability to probe multiple hadronic states with equal sensitivity; by contrast, Laplace sum-rules would suppress the contribution of the heavier state and obscure the multi-resonance content of Eqs.~(\ref{Pi_QCD_11},\ref{Pi_QCD_22}).  

Gaussian QCD sum-rules are founded upon quark-hadron duality, and relate a QCD prediction to hadronic physics via \cite{gauss}
\begin{gather}
G^{\rm QCD} \left({\hat s}, \tau\right) =G^{H} \left({\hat s}, \tau\right) \,,
\\
G^{H} \left({\hat s}, \tau\right) =\frac{1}{\sqrt{4\pi\tau}}
\int\limits_{t_0}^{\infty} \!\! dt \,{\rm exp} \left[  {\frac{-({\hat s} - t)^2}{4\tau}}\right]\,\trho^H(t)\,,
\label{GSR_hadronic}
\\
G^{\rm QCD}\left(\hat s,\tau\right)=\sqrt{\frac{\tau}{\pi}}\hat B\left[
\frac{\Pi^{\rm QCD}\left(-\hat s-i\Delta\right)-\Pi^{\rm QCD}\left(-\hat s+i\Delta\right)}{i\Delta}
\right]
\,,~
\hat B\equiv 
\lim_{\stackrel{N \rightarrow \infty~,~\Delta^2\rightarrow \infty}{\Delta^2/N\equiv 4\tau}}
\frac{\left(-\Delta^2\right)^N}{\Gamma(N)}\left(\frac{d}{d\Delta^2}\right)^N
\label{G_QCD}
\end{gather}
where $\trho^H(t)$ is the hadronic spectral function with threshold $t_0$
(note the distinction between the notation $\rho$ for spin projection factors of Eqs.~(\ref{E_rho_piK},\ref{E_rho_pieta}) and $\trho$ for hadronic spectral functions).   The quantity $\hat s$, corresponding to the peak of the Gaussian kernel, can be varied to scan the hadronic spectral function averaged over the approximate duality interval $ \hat s-2\sqrt{\tau}\lesssim t \lesssim \hat s+2\sqrt{\tau}$, but $\tau$ is constrained by QCD because it corresponds to the renormalization scale in light-quark systems \cite{harnett_quark}.  The hadronic spectral function is parameterized using the model 
\begin{equation}
\trho^H(t)=
\trho^{\rm res}(t) +\theta\left(t-s_0\right){\frac{1}{\pi}} {\rm Im} \Pi^{ \rm QCD}(t)\,,
\label{spectral_new}
\end{equation}
where $s_0$ is the continuum threshold inherent in QCD sum-rule methods \cite{SVZ,Reinders:1984sr,Narison:2002woh,Gubler:2018ctz,Colangelo:2000dp}
and $\trho^{\rm res}(t)$ is the hadronic resonance contribution to the spectral function.
Thus the Gaussian sum-rule relating a QCD prediction to the hadronic resonance(s) contribution is 
\begin{gather}
  {\cal G}^{\rm QCD} \left({\hat s}, \tau, s_0\right) =G^{\rm res} \left({\hat s}, \tau\right) \,,
\label{gsr_qcd_res}
\\
G^{\rm res} \left({\hat s}, \tau\right) =\frac{1}{\sqrt{4\pi\tau}}
\int\limits_{t_0}^{\infty} \!\! dt \,{\rm exp} \left[  {\frac{-({\hat s} - t)^2}{4\tau}}\right]\,\trho^{\rm res} (t)\,,  
\label{gsr_res}
\\
{\cal G}^{\rm QCD}\left({\hat s}, \tau, s_0\right)=
G^{\rm QCD} \left({\hat s}, \tau\right)-
\frac{1}{\sqrt{4\pi\tau}}
\int\limits_{s_0}^{\infty} \!\! dt \,{\rm exp} \left[  {\frac{-({\hat s} - t)^2}{4\tau}}\right]\,\frac{1}{\pi}{\rm Im}\Pi^{\rm QCD}(t)\,.  
\label{gsr_QCD_cont}
\end{gather}

Using the 2$\times$2 scale factor results of Eqs.~(\ref{Pi_QCD_11},\ref{Pi_QCD_22}) to guide a duality relation for the  
the Gaussian sum-rule expressions (\ref{gsr_qcd_res}--\ref{gsr_QCD_cont}) provides a Gaussian sum-rule framework relating QCD to chiral Lagrangians for the isodoublet/isotriplet sub-sectors
\begin{gather}
 \frac{m_q^2}{\Lambda^6}{\cal G}^{\rm QCD}_{(s)11}\left({\hat s}, \tau, s_0^{(1)}\right) =\cos^2{\theta_s}G_{(s)11}^{\rm res} \left({\hat s}, \tau\right)+  \sin^2{\theta_s}G_{(s)22}^{\rm res} \left({\hat s}, \tau\right)\,,
 \label{GSR_11}
 \\
 \frac{1}{\left(\Lambda'\right)^{10}}{\cal G}^{\rm QCD}_{(s)22}\left({\hat s}, \tau, s_0^{(2)}\right) =\sin^2{\theta_s}G_{(s)11}^{\rm res} \left({\hat s}, \tau\right)+  \cos^2{\theta_s}G_{(s)22}^{\rm res} \left({\hat s}, \tau\right)\,,
 \label{GSR_22}
\end{gather}
where the $s=\{\kappa,a\}$, $11$ and $22$ subscripts consequently follow from either the resonance contributions or the QCD correlation function, with a QCD continuum  
$\{s_0^{(1)},s_0^{(2)}\}$ in each case that naturally emerges from the corresponding QCD correlation function.
The $s=\{\kappa, a\}$ notation does not occur with  the scale factors to emphasize their universal nature. The mixing angles in Eqs.~(\ref{GSR_11},\ref{GSR_22}) as determined by Chiral Lagrangian analyses are \cite{GLSM} 
\begin{equation}
    \cos\theta_\kappa=0.4161\,,~\cos\theta_a=0.6304\,,
    \label{theta_values}
\end{equation}
as previously used in our benchmark scale-factor analysis \cite{CLQCDSR_2020}.

Eqs.~\eqref{GSR_11} and \eqref{GSR_22} represent a revised scale-factor methodology that can now be contrasted with Refs.~\cite{CLQCDSR_2016,CLQCDSR:2019_Proc,CLQCDSR_2020,Fariborz:2019zht}
by inverting (\ref{GSR_11},\ref{GSR_22}) to isolate the resonance contributions to find
{\allowdisplaybreaks
\begin{gather}
G_{(s)11}^{\rm res}(\hat s,\tau)=a A_s
{\cal G}_{(s)11}^{\rm QCD}\left(\hat s,\tau, s_0^{(1)}\right)-bB_s
{\cal G}_{(s)22}^{\rm QCD}\left(\hat s,\tau, s_0^{(2)}\right)
\label{G_eqs_1}
\\
G_{(s)22}^{\rm res}(\hat s,\tau)=-aB_s
{\cal G}_{(s)11}^{\rm QCD}\left(\hat s,\tau, s_0^{(1)}\right)+bA_s
{\cal G}_{(s)22}^{\rm QCD}\left(\hat s,\tau, s_0^{(2)}\right)
\label{G_eqs_2}
\\
A_s=\frac{\cos^2\theta_s}{\cos^2\theta_s-\sin^2\theta_s}\,,~
B_s=\frac{\sin^2\theta_s}{\cos^2\theta_s-\sin^2\theta_s}
\label{A_B_defn}
\\
a=\frac{m_q^2}{\Lambda^6}\,,~b=\frac{1}{\left(\Lambda'\right)^{10}}\,.
\end{gather}
Note that the $s=\{\kappa,a\}$ subscript is omitted from $a$ and $b$ because they contain the universal scale factors.  
By comparison of (\ref{G_eqs_1},\ref{G_eqs_2}) with e.g., Eq.~(19) of  Ref.~\cite{CLQCDSR_2020}, we see that the revised scale factor methodology results in a similar structure but entangles the QCD contributions (including the QCD continuum) in different ways because the  off-diagonal constraint \eqref{Pi_12_constraint} has not been used in the revised approach.\footnote{In Ref.~\cite{CLQCDSR_2020} the ``H" notation should be interpreted as ``res".} The expressions (\ref{G_eqs_1},\ref{G_eqs_2}) satisfy 
\begin{equation}
G_{(s)11}^{\rm res}(\hat s,\tau)+G_{(s)22}^{\rm res}(\hat s,\tau)
=a{\cal G}_{(s)11}^{\rm QCD}\left(\hat s,\tau, s_0^{(1)}\right)+b{\cal G}_{(s)22}^{\rm QCD}\left(\hat s,\tau, s_0^{(2)}\right)\,.
\label{trace_constraint}
\end{equation}
implying that they are a quark-hadron duality manifestation of the  trace relation  following from \eqref{phys_corr_new}  
(see Ref.~\cite{CLQCDSR_2020})
\begin{equation}
{\rm Tr}\left[ \Pi^P\right]
={\rm Tr}\left[ \Pi^{\rm QCD}I_s^2 \right]\,.
\label{trace_result}
\end{equation}
}

The universal scale factors can now be isolated from (\ref{G_eqs_1},\ref{G_eqs_2}) to find
\begin{gather}
\frac{1}{\Lambda^6}=\frac{A_s G_{(s)11}^{\rm res}(\hat s,\tau)+B_sG_{(s)22}^{\rm res}(\hat s,\tau)}{\left(A_s+B_s\right) m_q^2 \,{\cal G}_{(s)11}^{\rm QCD}\left(\hat s,\tau, s_0^{(1)}\right) }
\,,
\label{scale_relation_lambda}
\\[2pt]
\frac{1}{\left(\Lambda'\right)^{10}}=\frac{B_s G_{(s)11}^{\rm res}(\hat s,\tau)+A_sG_{(s)22}^{\rm res}(\hat s,\tau)}{\left(A_s+B_s\right) \, {\cal G}_{(s)22}^{\rm QCD}\left(\hat s,\tau, s_0^{(2)}\right) }
\,,
\label{scale_relation_lambda_prime}
\end{gather}
where the quantity $m_q^2$ is naturally combined into 
${\cal G}_{(s)11}^{\rm QCD}$ via renormalization group improvement as discussed below. Because the scale factors 
$\{\Lambda, \Lambda'\}$ are energy-independent, they are  determined by optimizing 
$\{s_0^{(1)},s_0^{(2)}\}$ to minimize the $\hat s$ dependence on the RHS of 
Eqs.~(\ref{scale_relation_lambda},\ref{scale_relation_lambda_prime}). This scale-factor determination  is done separately for each $s=\{\kappa,a\}$ channel as a test of universality. In the next Section we use (\ref{scale_relation_lambda},\ref{scale_relation_lambda_prime}) to analyze a variety of different resonance models and their impact on the predicted scale factors.

\section{Gaussian Sum-Rule  Resonance Models}
\label{models_section}

The expressions (\ref{scale_relation_lambda},\ref{scale_relation_lambda_prime}) relate the scale factors to Gaussian sum-rules containing  QCD predictions and  the resonance contributions.  We first outline the resonance models that will be studied.    The benchmark analysis of Refs.~\cite{CLQCDSR_2016,CLQCDSR:2019_Proc,CLQCDSR_2020,Fariborz:2019zht} was based upon 
\begin{equation}
\Pi^{\rm H}_{ij}  \left(q^2\right)
=i\int d^4x\, e^{iq\cdot x}
\langle 0| {\rm T} \left[ {\bf H}_i (x) {\bf H}_j(0) \right] |0 \rangle
=\delta_{ij} \,
\left(
\frac{1}
{m_{s i}^2 -q^2-i m_{s i}\Gamma_{si}}
\right)\,,~s=\{\kappa\,,a\}
\end{equation}
resulting in the the ``propagator" (PROP) resonance model 
\begin{gather}
\trho^{\rm res}_i= \trho_i^{\rm prop} (t)  
\\
\trho_i^{\rm prop} (t)  ={\rm Im}\Pi^H_{ii}(t)=\frac{m_i\Gamma_i}{\left(t-m_i^2\right)^2+m_i^2\Gamma_i^2}
\,,
\label{rho_prop}
\end{gather}
corresponding to a pure Breit-Wigner resonance shape. 
The $\Gamma_i\to 0$ limit provides an interesting special case for comparison, resulting in the narrow resonance (NR) model
\begin{equation}
  \trho_i^{\rm NR} (t) =\delta\left(t-m_i^2\right) \,.
\label{narrow_res}
\end{equation}
Although \eqref{narrow_res} would not be expected to be a reliable description of wide resonances such as the kappa $K_0^*(700)$, the NR model provides a test case to demonstrate that our Gaussian  sum-rule methodology is sensitive to width effects as discussed below.

Inspired by the structure of the complex resonance poles in the background-resonance interference approximation in Eqs.~(\ref{E_T012_rBW},\ref{E_T01_rBW}), we define the corresponding 
``extended distorted Breit-Wigner" (EDBW)
resonance model for a sum-rule analysis  
\begin{gather}
   \trho^{\rm res}_i(t) =\trho^{\rm EDBW}_i(t) 
   \\
 \trho^{\rm EDBW}_i(t)  =\frac{\rho(t)m_i\Gamma_i}{\left(t-m_i^2\right)^2+m_i^2\Gamma_i^2}+\xi_i \frac{\rho(t)\left(m_i^2-t\right)}{\left(t-m_i^2\right)^2+m_i^2\Gamma_i^2}\,,
    \label{rho_xi_fixed_width_new}
\end{gather}
where $\rho$ is defined for the isodoublet channel in 
Eq.~\eqref{E_rho_piK} and for the isotriplet channel in Eq.~\eqref{E_rho_pieta}. 
As discussed below, each resonance model is normalized to the benchmark PROP model, and therefore only a single parameter $\xi_i$ is necessary to describe the relative contributions of the real and imaginary parts of the complex resonance poles. 
The EDBW terminology is adopted because the limiting case $\xi_i=0$ leads to a Breit-Wigner structure with a distortion caused by the factor $\rho(t)$.    From the real part of 
\eqref{E_T012_rBW} and \eqref{E_T01_rBW} 
 we see that $\xi_i$ is identified with
\begin{equation}
    \xi^{(1)}_i=-\frac{A_{iR}}{A_{iI}}\,,
    \label{xi_v1_new}
\end{equation}
while from the imaginary part of \eqref{E_T012_rBW} and \eqref{E_T01_rBW}  $\xi_i$ is identified with
\begin{equation}
    \xi^{(2)}_i=\frac{A_{iI}}{A_{iR}}\,.
    \label{xi_v2_new}
\end{equation}
These two possible solutions  $
\xi_i=\{\xi^{(1)}_i, \xi^{(2)}_i \}$ will be considered in our analysis below.
Because the term proportional to $\xi_i$ in \eqref{rho_xi_fixed_width_new} is fitted in the region below $\approx 1.6\,{\rm GeV}$ (see Figs.~\ref{F_rBW}, \ref{F_rEBW}, \ref{F_pieta_fit_BW}, \ref{F_pieta_fit_EBW}) and also has poor convergence for large $t$, this term is cut-off above $t_c=(1.6\,{\rm GeV})^2$ using a rapidly-decaying Gaussian factor $\theta\left(t-t_c\right)\exp{\left[-\left(t-t_c\right)^2\right]}$ in GeV units. The case $\xi_i=0$ (corresponding to that considered in Ref.~\cite{GLSM_piK}) is denoted as the ``distorted Breit-Wigner"  (DBW) model  because it modifies the propagator model \eqref{rho_prop} with the spin projection factor $\rho(t)$ of Eqs.~(\ref{E_rho_piK},\ref{E_rho_pieta}). The DBW resonance shape thus provides an intermediate stage bridging the EDBW and PROP models.  

The structure of (\ref{E_T012_rEBW},\ref{E_T01_a0_EBW})  generalizes Eq.~\eqref{rho_xi_fixed_width_new} to provide the ``extended generalized Breit-Wigner" (EGBW) model 
\begin{gather}
\trho^{\rm EGBW}_i(t)  =\frac{m_i G_i(t)^2}{\left(t-m_i^2\right)^2+m_i^2G_i(t)^2}+\xi_i \frac{G_i(t)\left(m_i^2-t\right)}{\left(t-m_i^2\right)^2+m_i^2G_i(t)^2}\,,
    \label{rho_xi_energy_width_new}
    \\
G_i(t)=\frac{\rho(t)\Gamma_i}{\rho_{0i}}\,,
\label{G_i}
\end{gather}
where the same cut-off factor employed for \eqref{rho_xi_fixed_width_new} is also used for the term proportional to $\xi_i$ and we use the spin-appropriate definition for $\rho$ from Eqs.~(\ref{E_rho_piK},\ref{E_rho_pieta}).    From the  real and imaginary parts of (\ref{E_T012_rEBW},\ref{E_T01_a0_EBW})
we see that Eqs.~(\ref{xi_v1_new},\ref{xi_v2_new}) also apply to the EGBW model.  The $\xi_i=0$ case of \eqref{rho_xi_energy_width_new} is denoted as the ``generalized Breit-Wigner" (GBW) model because it generalizes the DBW  model to have an energy-dependent width factor governed by $G_i(t)$.  The GBW model  thus completes a sequence of increasingly sophisticated models: NR, PROP, DBW, GBW, EDBW, EGBW as summarized in Table~\ref{tab:my_label}. 

\begin{table}
    \centering
    \renewcommand{\arraystretch}{1.5}
    \begin{tabular}{|c|c|c|} \hline
        \textbf{Model Name (Equation Number)}  & \textbf{Abbreviation} & $\trho^{\rm res}_i(t)$ \\\hline
               Narrow Resonance \eqref{narrow_res} & NR & $\delta\left(t-m_i^2\right)$ \\\hline
        Propagator Resonance \eqref{rho_prop} & PROP & $\frac{m_i\Gamma_i}{\left(t-m_i^2\right)^2+m_i^2\Gamma_i^2}$ \\[5pt]\hline
        Distorted Breit-Wigner \eqref{rho_xi_fixed_width_new} $\left(\xi_i= 0\right)$ & DBW & $\frac{\rho(t)m_i\Gamma_i}{\left(t-m_i^2\right)^2+m_i^2\Gamma_i^2}$ \\[5pt]\hline
        Generalized  Breit-Wigner \eqref{rho_xi_energy_width_new} $\left(\xi_i= 0\right)$& GBW & $\frac{m_i G_i(t)^2}{\left(t-m_i^2\right)^2+m_i^2G_i(t)^2}$\\[5pt]\hline
        Extended Distorted Breit-Wigner \eqref{rho_xi_fixed_width_new}& EDBW & $\frac{\rho(t)m_i\Gamma_i}{\left(t-m_i^2\right)^2+m_i^2\Gamma_i^2}+\xi_i \frac{\rho(t)\left(m_i^2-t\right)}{\left(t-m_i^2\right)^2+m_i^2\Gamma_i^2}$\\[5pt]\hline
        Extended Generalized Breit-Wigner \eqref{rho_xi_energy_width_new}& EGBW & $\frac{m_i G_i(t)^2}{\left(t-m_i^2\right)^2+m_i^2G_i(t)^2}+\xi_i \frac{G_i(t)\left(m_i^2-t\right)}{\left(t-m_i^2\right)^2+m_i^2G_i(t)^2}$ \\[5pt]\hline
    \end{tabular}
    \caption{Summary of resonance models used in analysis. 
    Models are arranged in order of increasing sophistication and increasing width effects.}
    \label{tab:my_label}
\end{table}

One of our goals is to analyze the impact of different resonance models on the scale factors defined via Eqs.~(\ref{scale_relation_lambda},\ref{scale_relation_lambda_prime}).  
It is evident from Eqs.~(\ref{scale_relation_lambda},\ref{scale_relation_lambda_prime}) that changing the overall normalization of a resonance model will have an immediate impact on the scale factors, and thus adopting a common normalization for all resonance models is necessary for reliable comparisons between  them. Because the PROP model was used in our previous analysis in Refs.~\cite{CLQCDSR_2016,CLQCDSR:2019_Proc,CLQCDSR_2020,Fariborz:2019zht}, we choose the PROP model as the benchmark to establish the  normalization of the other models. 
As outlined in more detail below, the benchmark Gaussian sum-rule analysis of Refs.~\cite{CLQCDSR_2016,CLQCDSR:2019_Proc,CLQCDSR_2020,Fariborz:2019zht} was performed for $\tau=3\,{\rm GeV^4}$ in the region $0<\hat s<6\,{\rm GeV^2}$.  The normalization of each model is therefore defined such that
\begin{equation}
\int_0^{6\,{\rm GeV^2}} 
G_{ii}^{\rm res}\left(\hat s, \tau= 3\,{\rm GeV^4}\right)\,d\hat s=\int_0^{6\,{\rm GeV^2}} 
G_{ii}^{\rm prop}\left(\hat s, \tau= 3\,{\rm GeV^4}\right)\,d\hat s  \,.
\label{norm_constraint}
\end{equation}
The normalization constraint \eqref{norm_constraint} isolates the differences in the $\hat s$ dependence for each model within a common normalization, enabling a reliable and consistent comparison of scale factors across the full range  of resonance models.

\begin{figure}[hbt]
\centering
\includegraphics[scale=0.6]{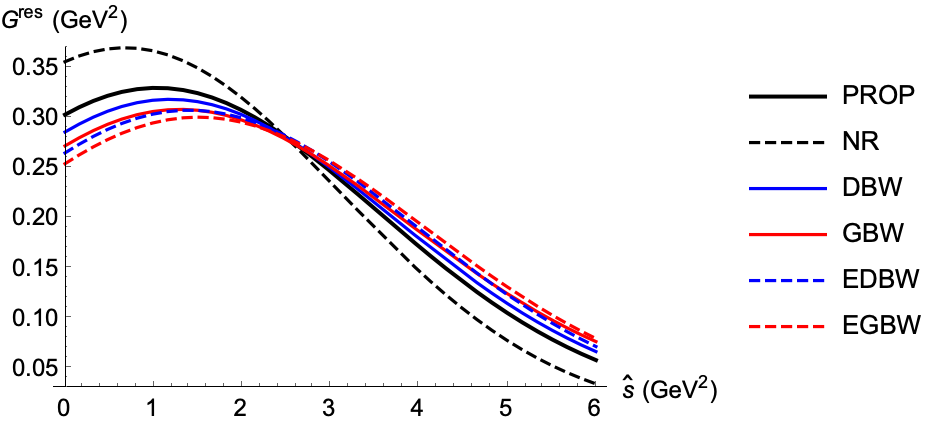}
    \caption{$G^{\rm res}$ [see \eqref{gsr_res}] for the $K_0^*(700)$ resonance models of Table~\ref{tab:my_label} plotted as a function of $\hat s$ for $\tau=3\,{\rm GeV^4}$ using the mass, width, and $\xi^{(1)}$ parameters of Table~\ref{gsr_res_parameter_tab}. All models are normalized via Eq.~\eqref{norm_constraint}.    }
    \label{kappa_model_fig}
\end{figure}

\begin{figure}[hbt]
\centering
\includegraphics[scale=0.6]{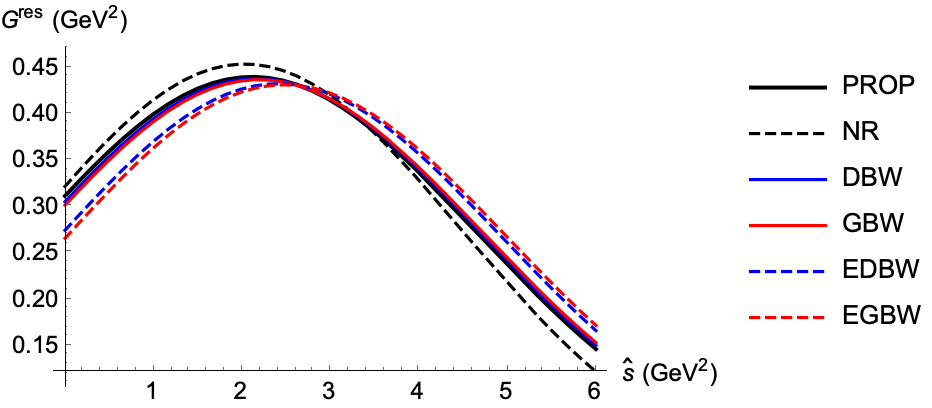}
    \caption{$G^{\rm res}$ [see \eqref{gsr_res}] for the $K_0^*(1430)$ resonance models of Table~\ref{tab:my_label} plotted as a function of $\hat s$ for $\tau=3\,{\rm GeV^4}$ using the mass, width, and $\xi^{(1)}$ parameters of Table~\ref{gsr_res_parameter_tab}. All models are normalized via Eq.~\eqref{norm_constraint}. }
    \label{k01430_model_fig}
\end{figure}

\begin{figure}[hbt]
\centering
\includegraphics[scale=0.6]{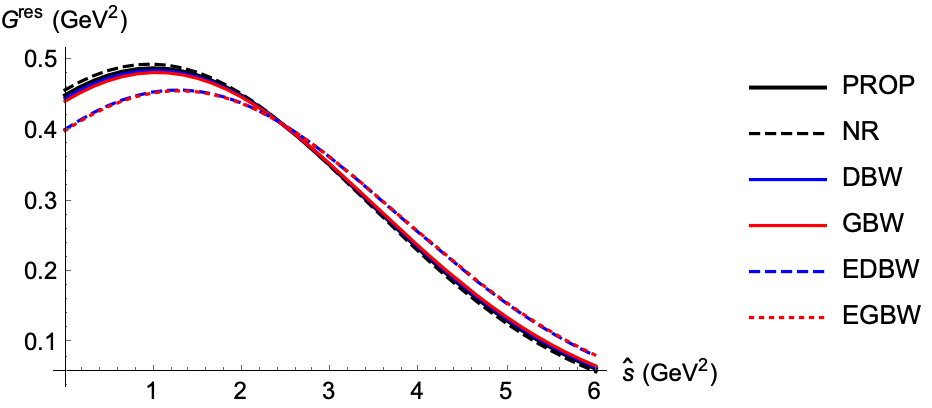}
    \caption{$G^{\rm res}$ [see \eqref{gsr_res}] for the $a_0{(980)}$ resonance models of Table~\ref{tab:my_label} plotted as a function of $\hat s$ for $\tau=3\,{\rm GeV^4}$ using the mass, width, and $\xi^{(2)}$ parameters of Table~\ref{gsr_res_parameter_tab}. All models are normalized via Eq.~\eqref{norm_constraint}. }
    \label{a0_model_fig}
\end{figure}

\begin{figure}[hbt]
\centering
\includegraphics[scale=0.6]{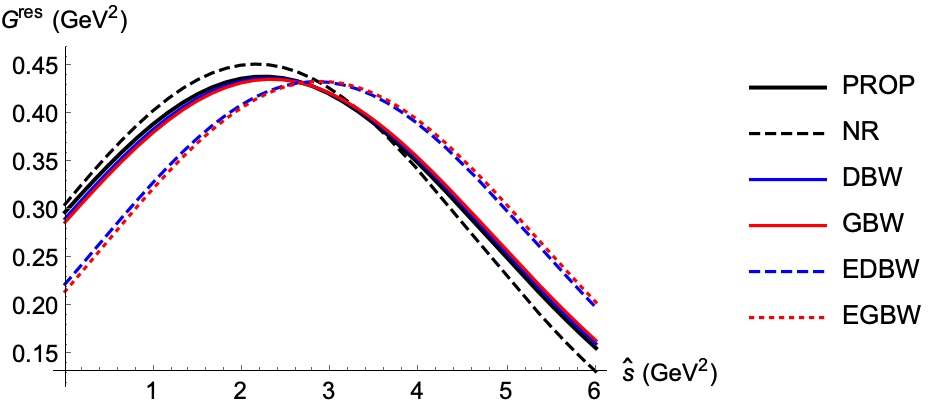}
    \caption{$G^{\rm res}$ [see \eqref{gsr_res}] for the $a_0{(1450)}$ resonance models of Table~\ref{tab:my_label} plotted as a function of $\hat s$ for $\tau=3\,{\rm GeV^4}$ using the mass, width, and $\xi^{(2)}$ parameters of Table~\ref{gsr_res_parameter_tab}. All models are normalized via Eq.~\eqref{norm_constraint}.}
    \label{a0_prime_model_fig}
\end{figure}

The ability of the Gaussian sum-rules to distinguish between the different resonance models can now be examined.  In particular, Eq.~\eqref{gsr_res} is calculated for each resonance model using the parameters of Table~\ref{gsr_res_parameter_tab} (corresponding to the benchmark analysis of \cite{CLQCDSR_2020}).  Figures \ref{kappa_model_fig}--\ref{a0_prime_model_fig} compare $G^{\rm res}$ in Eq.~\eqref{gsr_res} for the models of Table~\ref{tab:my_label} using the Table~\ref{gsr_res_parameter_tab} parameters and the normalization constraint of  Eq.~\eqref{norm_constraint}. 

\begin{table}[htb]
\centering
\renewcommand{\arraystretch}{1.5}
\begin{tabular}{|c|c|c|c|c|c|c|}
\hline
Resonance & Mass (GeV) & Width (GeV) & $\xi^{(1)}$ (EDBW) & $\xi^{(2)}$ (EDBW) & $\xi^{(1)}$ (EGBW)   & $\xi^{(2)}$ (EGBW)
\\
\hline
$K_0^*(700)$ & $0.824$ & $0.478$ & $-0.3889$ & $2.571$ 
& $-0.4283$ & $2.335$
\\
\hline
$K_0^*(1430)$ & $1.425$ & $0.270$ & $-0.3727$ & $2.683$ 
& $-0.4244$ & $2.356$
\\
\hline
$a_0(980)$ & $0.980$ & $0.06$ &  $2.899$  &   $-0.3449$ 
&  $3.027$  &  $-0.3303$  
\\
\hline
$a_0(1450)$ & $1.47$ & $0.265$ &  $1.132$  &   $-0.8835$ 
&  $1.119$  &  $-0.8937$ 
\\
\hline
\end{tabular}
\caption{
Parameters  for the resonance models summarized in Table~\ref{tab:my_label}.  The parameters $\{\xi^{(1)}_i,\xi^{(2)}_i\}$ are obtained from 
(\ref{xi_v1_new},\ref{xi_v2_new}) combined with Tables~\ref{T_BW_EBW_fits} and \ref{T_BW_EBW_a0_fits}. The masses and widths correspond to our benchmark analysis \cite{CLQCDSR_2020} and are consistent with the PDG ranges \cite{PDG}.}
\label{gsr_res_parameter_tab}
\end{table}

Figures \ref{kappa_model_fig}--\ref{a0_prime_model_fig} demonstrate that the Gaussian sum-rule can distinguish between the different models, with the $K_0^*(700)$ showing a  particularly clear distinction between the various models because of its large width.  An important aspect of these Figures is the overall sensitivity of the Gaussian sum-rules to width effects as illustrated by separation of the narrow-width NR case from models that incorporate resonance widths.  The only exception is for the rather narrow $a_0(980)$ state, but even in this situation the EDBW and EGBW cases are clearly resolved from the other models.  This demonstrated  ability of the Gaussian sum-rules to resolve width effects and distinguish between resonance models, particularly for light and broad resonances such as the kappa,  is one of the key findings in this work. 
As outlined below, this distinction between resonance models is quantified by measures of scale factor universality and energy-independence.

\begin{figure}[hbt]
\centering
\includegraphics[scale=0.6]{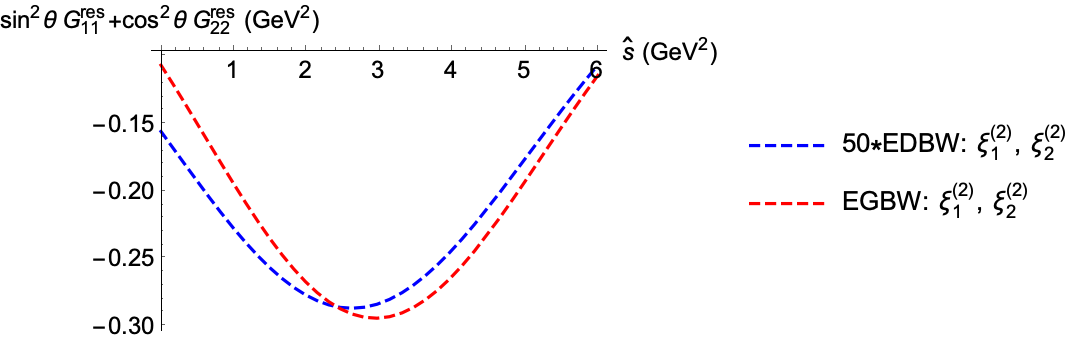}
    \caption{The RHS of  \eqref{GSR_11}  for the EDBW and EGBW models for the isodoublet system  is plotted as a function of $\hat s$ for $\tau=3\,{\rm GeV^4}$ using the mass, width, and $\xi^{(2)}$ parameters of Table~\ref{gsr_res_parameter_tab} and the mixing angle of Eq.~\eqref{theta_values}. The  normalization constraint \eqref{norm_constraint} has not been employed. }
    \label{isodoublet_alt_solution_fig}
\end{figure}

\begin{figure}[hbt]
\centering
\includegraphics[scale=0.6]{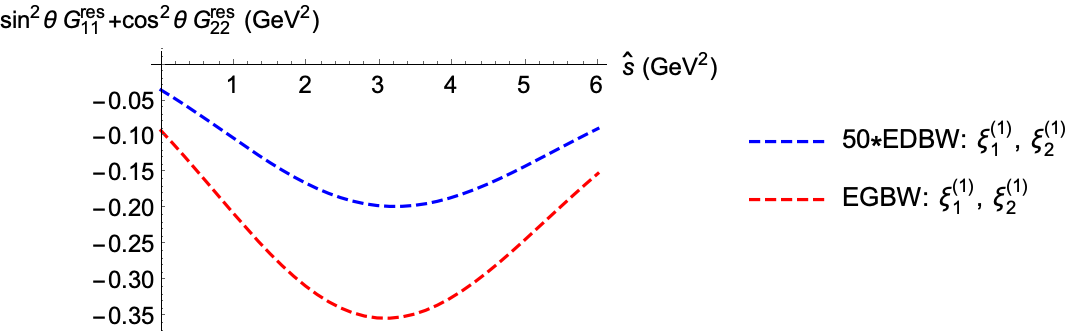}
    \caption{The RHS of  \eqref{GSR_11}  for the EDBW and EGBW models for the isotriplet system  is plotted as a function of $\hat s$ for $\tau=3\,{\rm GeV^4}$ using the mass, width, and $\xi^{(1)}$ parameters of Table~\ref{gsr_res_parameter_tab} and the mixing angle of Eq.~\eqref{theta_values}. The  normalization constraint \eqref{norm_constraint} has not been employed.}
       \label{isotriplet_alt_solution_fig}
\end{figure}

In forming Figures \ref{kappa_model_fig}--\ref{a0_prime_model_fig}, one of the two possible $\xi^{(j)}$ values from Table~\ref{gsr_res_parameter_tab} has been selected.  However, the Gaussian sum-rules can be used to demonstrate that the other solution is non-physical. In Eqs.~(\ref{GSR_11},\ref{GSR_22}), the QCD side of the equation is manifestly positive, and therefore the resonance contributions on the RHS must  also be positive for the existence of a scale-factor bridge between QCD and Chiral Lagrangians.  Using the mixing angles of Eq.~\eqref{theta_values}, Figs.~\ref{isodoublet_alt_solution_fig}--\ref{isotriplet_alt_solution_fig} show that these alternative solutions lead to a negative RHS of \eqref{GSR_22}, and hence the associated $\xi^{(j)}$ solutions are non-physical and will be omitted from the detailed scale-factor analysis of Section~\ref{analysis_section} below.

\section{Gaussian QCD Sum-Rule Analysis of Scale Factors}
\label{analysis_section} 
The QCD contributions to the Gaussian sum-rules \eqref{gsr_QCD_cont} are needed to analyze the scale factor relation connecting QCD to  Chiral Lagrangians in Eqs.~(\ref{scale_relation_lambda},\ref{scale_relation_lambda_prime}). The correlation function for  the two-quark current $J_1^\kappa$ \eqref{J1_kappa} is given in \cite{Zhang:2009qb,Du:2004ki,Jamin:1992se,Reinders:1984sr} and the methods of \cite{harnett_quark} can  be used to form the QCD Gaussian sum-rule 
{\allowdisplaybreaks
\begin{gather}
\begin{split}
&{\cal G}_{(\kappa) 11}^{\rm QCD}\left(\hat s, \tau,s_0\right)=
\frac{3}{8\pi^2}
\int\limits_0^{s_0} \!\!
t\,dt\left[\left(1+\frac{17}{3}\frac{\alpha_s}{\pi}\right)
-2\frac{\alpha_s}{\pi}\log{\left(\frac{t}{\sqrt{\tau}}\right)}
\right]W\left(t,\hat s,\tau\right)
\\
&  \phantom{{\cal G}} +\frac{\pi n_c\rho_c^2 }{m_s^*m_q^*}
\int\limits_0^{s_0}
t J_1\left(\rho_c\sqrt{t}\right) Y_1\left(\rho_c\sqrt{t}\right) 
W\left(t,\hat s,\tau\right)\,dt
+ \exp{\left(-\frac{\hat s^2}{4\tau}\right)}\left[
\frac{1}{2\sqrt{\pi\tau}}\left\langle C^\kappa_4{\cal O}^\kappa_4\right\rangle-
\frac{\hat s}{4\tau\sqrt{\pi\tau}}\left\langle C^\kappa_6{\cal O}^\kappa_6\right\rangle
\right]~,
\end{split}
\label{gauss_scalar_QCD}
\\
W\left(t,\hat s,\tau\right)=\frac{1}{\sqrt{4\pi\tau}}\exp{\left(-\frac{\left(t-\hat s\right)^2}{4\tau}\right)}
\\
\left\langle C_4^s {\cal O}_4^\kappa\right\rangle=
 \left\langle m_s \overline{q}q\right\rangle 
+\frac{1}{2} \left\langle m_s \overline{s}s\right\rangle +
\frac{1}{8\pi} \left\langle\alpha_s G^2 \right\rangle\,,~
\label{c4_scalar}
\\
\langle C^\kappa_6{\cal O}^\kappa_6\rangle=
-\frac{1}{2}\left\langle m_s\overline{q}\sigma G q\right\rangle
-\frac{1}{2}\left\langle m_q\overline{s}\sigma G s\right\rangle
-\frac{16\pi}{27}\alpha_s
\left( \left\langle \bar q q\right\rangle^2+\left\langle \bar s s\right\rangle^2  \right)
-\frac{48}{9}\alpha_s\left\langle \bar q q\right\rangle\left\langle \bar s s\right\rangle
\,,
\label{c6_scalar}
\end{gather}
}where $q$ denotes the non-strange ($u,d$) quarks,  vacuum saturation has been used  for the dimension-six (four-quark) condensates, and definitions of other parameters will be specified below.
Note that $Y_1$ and $J_1$ in \eqref{gauss_scalar_QCD} represent Bessel functions; the similar notation $J_1^\kappa$ in \eqref{J1_kappa} and $J_1^a$ in \eqref{J_a_QCD_new}   have an additional subscript to identify them as currents. Similarly, the quantity $\rho_c$ is an 
instanton-liquid parameter \cite{Shuryak:1982qx} and should not be confused with the spin projection factors in Eqs.~(\ref{E_rho_piK},\ref{E_rho_pieta}). 
The combination $m_q^2 {\cal G}_{(\kappa) 11}$ occurring in \eqref{scale_relation_lambda}
implies that this combined quantity satisfies a homogeneous renormalization-group equation, and hence 
all running quantities are specified 
at the renormalization scale $\nu^2=\sqrt{\tau}$ \cite{harnett_quark}. 

Similarly, the Gaussian sum-rule related  to the four-quark current 
$J^\kappa_2$ \eqref{J2_kappa}
is obtained via  \cite{Chen:2007xr,Chen:2006hyp} 
\begin{gather}
{\cal G}_{(\kappa)22}^{\rm QCD}\left(\hat s, \tau,s_0\right)=
\int\limits_{0}^{s_0}\!\!dt\,   W\left(t,\hat s,\tau\right)
\, \rho_{(\kappa)}^{\rm QCD}(t) \,,
\label{G_kappa_22}
\\
\begin{split}
\rho_{(\kappa)}^{\rm QCD}(t)=&
\left(\frac{t^4}{11520 \pi ^6}-\frac{t^3m_s^2 }{572 \pi ^6}\right)
+
t^2 \left(\frac{\left(7+6 \sqrt{2}\right) \left\langle\alpha_s G^2 \right\rangle}{2304 \pi ^5}+\frac{\left\langle m_s \overline{s}s\right\rangle}{72
   \pi ^4}\right)
  \\
  &+t \left(\frac{\left(-7-6 \sqrt{2}\right) \aGG m_s^2}{768 \pi
   ^5}+\frac{ \msqGq}{128 \pi ^4}\right)
\\
&-\frac{\aGG \msqq}{96 \pi ^3}+\frac{\left(7+6 \sqrt{2}\right)
   \aGG \mss}{576 \pi ^3}-\frac{ \qGq \ss}{48 \pi
   ^2}+\frac{ \qq \sGs}{48 \pi ^2}\,,
\end{split}   
 \label{rho_kappa_4q}
\end{gather}
where vacuum saturation has been used  for the dimension-eight  condensates.
Renormalization effects associated with the current $J_2^\kappa$ represents a higher-order effect, so 
the leading-order  result \eqref{rho_kappa_4q}  and the Gaussian sum-rule $G_{22}^{\rm QCD}$ effectively satisfies a homogeneous renormalization-group equation, allowing application of the renormalization-group results of Ref.~\cite{harnett_quark}.

Proceeding in a similar way for the isotriplet sector, the correlation 
function for  the two-quark current $J_1^a$ \eqref{J_a_QCD_new}
can be found in Refs.~\cite{SVZ,Chetyrkin:1996sr,Gorishnii:1991zr,Gorishnii:1990zu,Bagan:1985zp,Reinders:1981ww,Shuryak:1982qx}
and the resulting Gaussian sum-rule is given in \cite{harnett_quark}
\begin{gather}
\begin{split}
&{\cal G}_{(a) 11}^{\rm QCD}\left(\hat s, \tau,s_0\right)=
\frac{3}{8\pi^2}
\int\limits_0^{s_0} \!\!
t\,dt\left[\left(1+\frac{17}{3}\frac{\alpha_s}{\pi}\right)
-2\frac{\alpha_s}{\pi}\log{\left(\frac{t}{\sqrt{\tau}}\right)}
\right]W\left(t,\hat s,\tau\right)
\\
&  \phantom{{\cal G}} +
\frac{3}{4\pi}
\int\limits_0^{s_0}
t J_1\left(\rho_c\sqrt{t}\right) Y_1\left(\rho_c\sqrt{t}\right) 
W\left(t,\hat s,\tau\right)\,dt
+ \exp{\left(-\frac{\hat s^2}{4\tau}\right)}\left[
\frac{1}{2\sqrt{\pi\tau}}\left\langle C^a_4{\cal O}^a_4\right\rangle-
\frac{\hat s}{4\tau\sqrt{\pi\tau}}\left\langle C^a_6{\cal O}^a_6\right\rangle
\right]~,
\end{split}
\label{gauss_scalar_QCD_a0}
\\
\left\langle C_4^s {\cal O}_4^a\right\rangle=
3\mqq+\frac{1}{8\pi}\aGG
\,,~
\langle C^a_6{\cal O}^a_6\rangle=
-\frac{172}{27}\alpha_s\qq^2
\,.
\label{c4_c6_scalar_a0}
\end{gather}
As for the $\kappa$ case,  the combination $m_q^2 {\cal G}_{(a) 11}$ occurring in \eqref{scale_relation_lambda}
implies that  renormalization-group effects for this combined quantity 
are specified 
at the renormalization scale $\nu^2=\sqrt{\tau}$ \cite{harnett_quark}, and vacuum saturation has been used  for the dimension-six (four-quark) condensates.

Finally, the Gaussian sum-rule related  to the four-quark current $J^a_2$ \eqref{J2_a}
is obtained via  \cite{Chen:2007xr,Chen:2006hyp}
\begin{gather}
{\cal G}_{(a)22}^{\rm QCD}\left(\hat s, \tau,s_0\right)=
\int\limits_{0}^{s_0}\!\!dt\,   W\left(t,\hat s,\tau\right)
\, \rho_{(a)}^{\rm QCD}(t) \,,
\label{G_a_22}
\\
\begin{split}
\rho_{(a)}^{\rm QCD}(t)=&
\frac{t^4}{11520 \pi ^6}-\frac{t^3 m_s^2}{288 \pi ^6}
+
t^2 \left(\frac{\left(7+6 \sqrt{2}\right) \aGG}{2304 \pi ^5}+\frac{\mss}{36
   \pi ^4}\right)
   \\
   &+t \left(\frac{\left(-7-6 \sqrt{2}\right) \aGG m_s^2}{384 \pi
   ^5}-\frac{\mss m_s^2}{6 \pi ^4}\right)
   \\
   &
  -\frac{\aGG \msqq }{48 \pi ^3}+\frac{\left(7+6 \sqrt{2}\right)
   \aGG \mss}{288 \pi ^3}+\frac{4 \msqq^2}{9 \pi
   ^2}+\frac{4 \mss^2}{9 \pi ^2}\,,
\end{split}   
 \label{rho_a0_4q}
\end{gather}
where vacuum saturation has been used  for the dimension-eight condensates.
As for the $\kappa$ case, ${\cal G}_{(a)22}^{\rm QCD}$ effectively satisfies a homogeneous renormalization-group equation, allowing application of the renormalization-group results of Ref.~\cite{harnett_quark}.

The QCD input parameters needed for the Gaussian sum-rules 
in Eqs.~(\ref{gauss_scalar_QCD},\ref{G_kappa_22},\ref{gauss_scalar_QCD_a0},\ref{G_a_22}), and as used in our benchmark analysis \cite{CLQCDSR_2020},
include values within the PDG ranges  for the quark masses and $\alpha_s$ \cite{PDG} along with  
the following QCD condensate \cite{Reinders:1984sr,Narison:2011rn,Beneke:1992ba,Belyaev:1982sa} and instanton-liquid model parameters 
\cite{Shuryak:1982qx,Schafer:1996wv}
{\allowdisplaybreaks
\begin{gather}
 \langle\alpha_s G^2\rangle
      = (0.07\pm 0.02)\, {\rm GeV^4} \,,
      \label{GG}
      \\
\frac{\left\langle \overline{q}\sigma G q\right\rangle}{\langle \bar q q\rangle}=\frac{\left\langle \overline{s}\sigma G s\right\rangle}{\langle \bar s s\rangle}=(0.8\pm 0.1) \,{\rm GeV^2}
\label{mix}
\\
\langle \bar q q\rangle=-\left(0.24\pm 0.2 \,{\rm GeV}\right)^3\,,~\langle \bar s s\rangle=(0.8\pm 0.1)\langle \bar q q\rangle
\label{O6}
\\
2\mqq=-f_\pi^2m_\pi^2\,, f_\pi = 130/\sqrt{2} {\rm MeV}
\\
  n_{{c}} = 8.0\times 10^{-4}\ {\rm GeV^4}~,~\rho_c =1/600\,{\rm MeV}~,
  \label{inst}
  \\
  m^*_q=170\,{\rm MeV}\,,~m^*_s=220\,{\rm MeV}~.
\end{gather}
} 
The instanton parameters $\rho_c$ and $n_c$ have an estimated uncertainty of 15\% and the quark zero-mode effective masses $m^*$ are correlated with the uncertainty in $\rho_c$ and the quark condensate \cite{Shuryak:1981ff}. 
The Ref.~\cite{PDG}  value used for the quark mass ratio $m_s/m_q=27.3$ is of particular importance because it appears in both the QCD inputs and as a parameter in the chiral Lagrangian analysis used to determine the Eq.~\eqref{theta_values} mixing angles.\footnote{The theoretical uncertainty in the gluon condensate (the dominant QCD condensate effect) on the $a_0$ scale factors was examined in Ref.~\cite{Fariborz:2019zht} and found to be a small numerical effect.} As mentioned previously, we use  $\tau=3\,{\rm GeV^4}$ consistent with our benchmark scale factor analysis \cite{CLQCDSR_2020} and other applications of  QCD Gaussian sum-rules in Refs.~\cite{harnett_quark,Harnett:2000fy}.

Returning to Eqs.~\eqref{scale_relation_lambda} and \eqref{scale_relation_lambda_prime}, for each channel the scale factors are naturally expressed as
\begin{gather}
 \lambda_s\left(\hat s,\tau,s_0^{(1)}\right)=  \Lambda
\,,~~~  
\lambda_s\left(\hat s,\tau,s_0^{(1)}\right)=  \left[\frac{\left(A_s+B_s\right) m_q^2 \,{\cal G}_{(s)11}^{\rm QCD}\left(\hat s,\tau, s_0^{(1)}\right) }{A_s G_{(s)11}^{\rm res}(\hat s,\tau)+B_sG_{(s)22}^{\rm res}(\hat s,\tau)}\right]^{1/6}\,,
  \label{Lambda_expression}
  \\
 \lambda'_s\left(\hat s,\tau,s_0^{(2)}\right)= \Lambda'\,,~~~
 \lambda'_s\left(\hat s,\tau,s_0^{(2)}\right)=\left[\frac{\left(A_s+B_s\right) \, {\cal G}_{(s)22}^{\rm QCD}\left(\hat s,\tau, s_0^{(2)}\right) }{B_s G_{(s)11}^{\rm res}(\hat s,\tau)+A_sG_{(s)22}^{\rm res}(\hat s,\tau)}\right]^{1/10}\,.
  \label{Lambda_prime_expression}
\end{gather}
Eqs.~\eqref{Lambda_expression} and \eqref{Lambda_prime_expression} draw together multiple elements including the QCD predictions ${\cal G}_{(s)ii}^{\rm QCD}$, Chiral Lagrangian mixing angles embedded in $\{A_s,B_s\}$ [see Eq.~\eqref{A_B_defn}], resonance models as inspired by the Section~\ref{res_shape_section} Chiral Lagrangian analysis of $\pi K$ and $\pi\eta$ scattering, and the scale factor bridge  connecting QCD to Chiral Lagrangian elements.   
For a particular resonance model and choice of $s_0^{(1)}$ and $s_0^{(2)}$, the scale factors in Eqs.~\eqref{Lambda_expression} and  \eqref{Lambda_prime_expression} are fitted to $\{\lambda_s,\lambda'_s\}$ over equally-spaced $\hat s$ values (usually about 25 points) over the range $0<\hat s<6\,{\rm GeV^2}$ and $\tau=3\,{\rm GeV^4}$ as used in our benchmark analysis \cite{CLQCDSR_2020}, leading to the fitted scale factors as a function of the continuum parameters
\begin{gather}
    \Lambda_{\rm fit}\left(s_0^{(1)}\right)=  
    \frac{\sum\limits_{\hat s} 
    \lambda_s\left(\hat s,\tau,s_0^{(1)}\right)
    }{ \sum\limits_{\hat s} 1}
    \,,
    \label{Lambda_fit}
    \\
  \Lambda'_{\rm fit}\left(s_0^{(2)}\right)= 
  \frac{\sum\limits_{\hat s}
  \lambda'_s\left(\hat s,\tau,s_0^{(2)}\right)
  }{ \sum\limits_{\hat s} 1}\,.
  \label{Lambda_prime_fit}
\end{gather}
The continuum parameters are then optimized by minimizing the following residual sum of squares that quantify scale-factor energy-independence
\begin{gather}
\chi^2_\Lambda \left(s_0^{(1)}\right)=\sum\limits_{\hat s}
\left(\frac{ \Lambda_{\rm fit}\left(s_0^{(1)}\right)}
 {\lambda_s\left(\hat s,\tau,s_0^{(1)}\right)}-1\right)^2\,,
\label{chi2_Lambda}
\\
\chi^2_{\Lambda'}\left(s_0^{(2)}\right)=\sum\limits_{\hat s}
\left(
\frac{ \Lambda'_{\rm fit}\left(s_0^{(2)}\right)}
{\lambda'_s\left(\hat s,\tau,s_0^{(2)}\right)}
-1
\right)^2
\,.
\label{chi2_Lambda_prime}
\end{gather}
An additional physical constraint for a valid optimization is that the continuum parameters must be larger than the masses in the channel. 
Within the chosen resonance model, this optimization procedure then results in the scale factors 
$\Lambda_{s }$ and  $\Lambda'_{s }$ 
for each channel, which can then be tested for universality  using the following measure 
\begin{equation}
    \Delta=\left\vert 
   \frac{ \Lambda_\kappa-\Lambda_a}{ \Lambda_\kappa+\Lambda_a}
    \right \vert 
    +
  \left\vert 
   \frac{ \Lambda'_\kappa-\Lambda'_a}{ \Lambda'_\kappa+\Lambda'_a}
    \right \vert   \,.
    \label{Delta_def}
\end{equation}

\begin{figure}[hbt]
\centering
\includegraphics[scale=0.55]{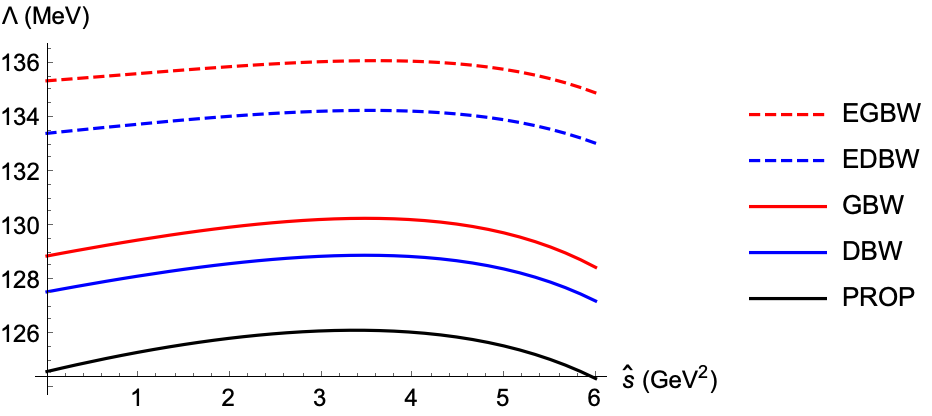}
\includegraphics[scale=0.55]{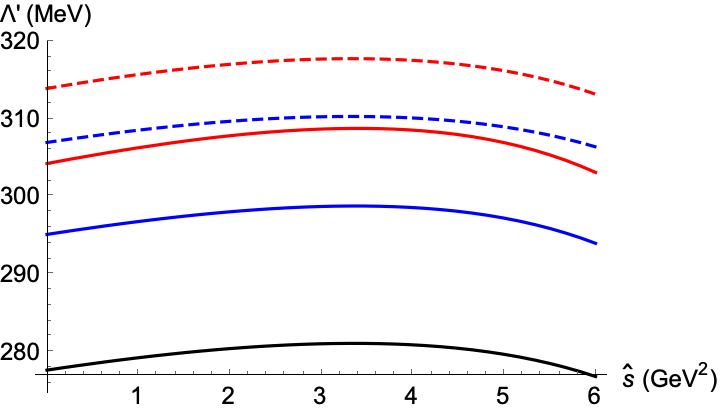}
    \caption{Theoretical predictions $\{\lambda_\kappa, \lambda'_\kappa\}$  for the scale factors [see Eqs.~\eqref{Lambda_expression} and \eqref{Lambda_prime_expression}] are shown as a function of $\hat s$
    in the isodoublet channel for the models and continuum values in Table~\ref{scale_factor_table}. The benchmark analysis value $\tau=3\,{\rm GeV^4}$ and central values of the QCD parameters have been used. 
    Scale has been chosen to highlight the differences between the models.
    }     
    \label{isodoublet_scale_factor_models_fig}
\end{figure}

\begin{figure}[hbt]
\centering
\includegraphics[scale=0.55]{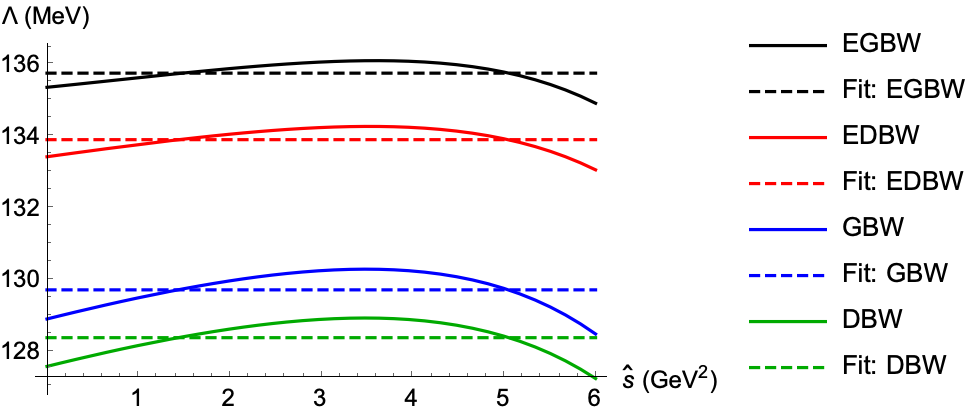}
\includegraphics[scale=0.55]{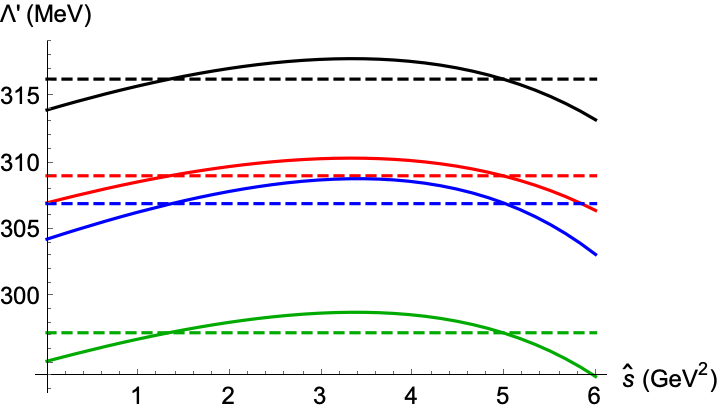}
    \caption{Fitted values of the scale factors [see Eqs.~\eqref{Lambda_fit} and \eqref{Lambda_prime_fit}] are compared with  the theoretical predictions $\{\lambda_\kappa, \lambda'_\kappa\}$   [see Eqs.~\eqref{Lambda_expression} and \eqref{Lambda_prime_expression}]  as a function of $\hat s$
    in the isodoublet channel for selected models and continuum values in Table~\ref{scale_factor_table}. The benchmark analysis value $\tau=3\,{\rm GeV^4}$ and central values of the QCD parameters have been used. 
    Scale has been chosen to highlight the differences between the models.
    }
\label{isodoublet_scale_factor_fits_fig}
\end{figure}

\begin{figure}[hbt]
\centering
\includegraphics[scale=0.55]{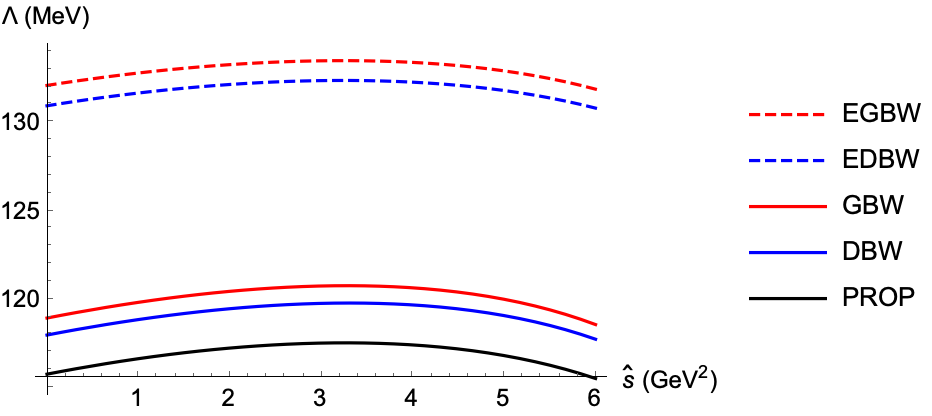}
\includegraphics[scale=0.55]{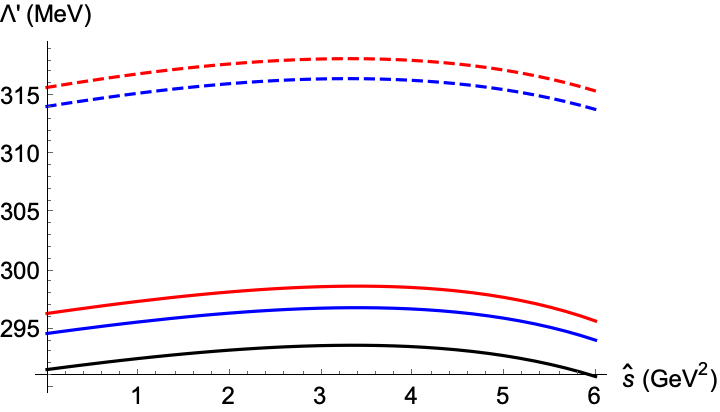}
    \caption{Theoretical predictions $\{\lambda_a, \lambda'_a\}$  for the scale factors [see Eqs.~\eqref{Lambda_expression} and \eqref{Lambda_prime_expression}] are shown as a function of $\hat s$
    in the isotriplet channel for the models and continuum values in Table~\ref{scale_factor_table}. The benchmark analysis value $\tau=3\,{\rm GeV^4}$ and central values of the QCD parameters have been used. 
    Scale has been chosen to highlight the differences between the models.
    }  \label{isotriplet_scale_factor_models_fig}
\end{figure}

\begin{figure}[hbt]
\centering
\includegraphics[scale=0.55]{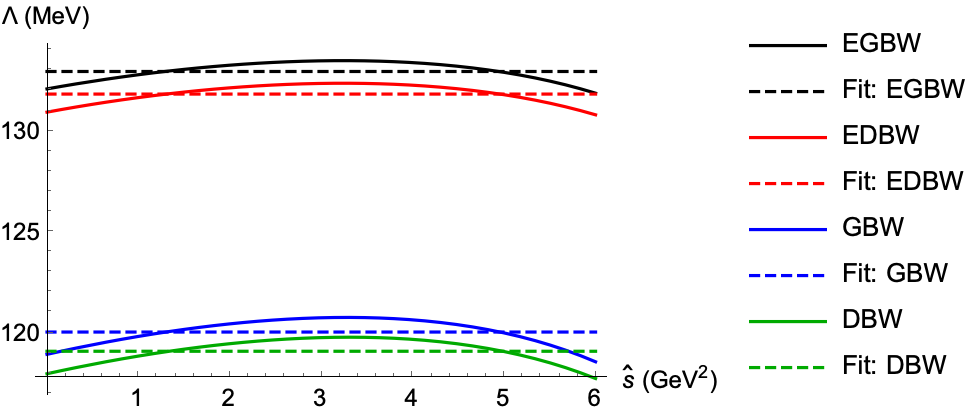}
\includegraphics[scale=0.55]{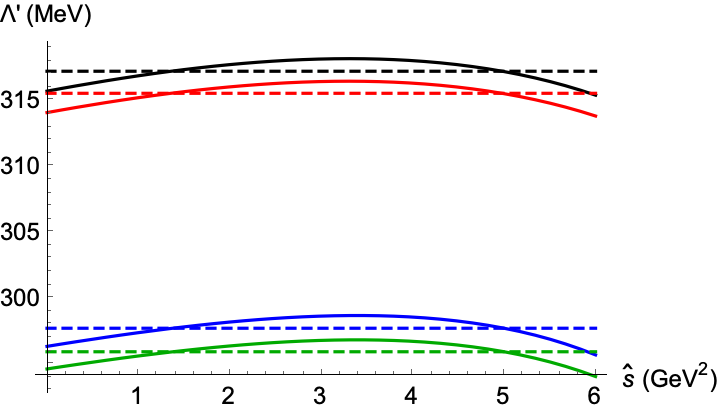}
    \caption{Fitted values of the scale factors [see Eqs.~\eqref{Lambda_fit} and \eqref{Lambda_prime_fit}] are compared with  the theoretical predictions $\{\lambda_a, \lambda'_a\}$   [see Eqs.~\eqref{Lambda_expression} and \eqref{Lambda_prime_expression}]  as a function of $\hat s$
    in the isotriplet channel for selected models and continuum values in Table~\ref{scale_factor_table}. The benchmark analysis value $\tau=3\,{\rm GeV^4}$ and central values of the QCD parameters have been used.
    Scale has been chosen to highlight the differences between the models.
    }
\label{isotriplet_scale_factor_fits_fig}
\end{figure}

The results of the optimization procedure are given in Table~\ref{scale_factor_table} for each channel and model, using the benchmark analysis value $\tau=3\,{\rm GeV^4}$ and central values of the QCD parameters. 
Figures~\ref{isodoublet_scale_factor_models_fig}--\ref{isotriplet_scale_factor_fits_fig} 
illustrate the theoretical predictions $\{\lambda_s, \lambda'_s\}$ [see Eqs.~\eqref{Lambda_expression} and \eqref{Lambda_prime_expression}]  and the fitted values of the scale factors  [see Eqs.~\eqref{Lambda_fit} and \eqref{Lambda_prime_fit}] for the corresponding models and channels.

\begin{table}[htb]
\centering
\renewcommand{\arraystretch}{1.5}
\begin{tabular}{|c|c|c|c|c|c|c|c|c|}
\hline
Model & Channel &   $s_0^{(1)}$ & $s_0^{(2)}$ & $\Lambda$ & $\Lambda'$  & $\chi^2_\Lambda\times 10^6$ & $\chi^2_{\Lambda'} \times 10^6$ & $\Delta$ \\
\hline
\hline
\multirow{2}{*}{PROP} 
& $K_0^*$ & $2.76$ & $1.79$ & $0.1256$ & $0.2798$ 
& $17.9$
&  $19.6$ 
& \multirow{2}{*}{$0.0586$}
\\
\hhline{|~|-|-|-|-|-|-|-|~|}
  & $a_0$  & $2.40$ & $2.13$ & $0.1169$ & $0.2928$  
 & $27.1$
 &  $6.83$  
 &
 \\
 \hline\hline
 \multirow{2}{*}{DBW} 
& $K_0^*$ & $2.90$ & $2.11$ & $0.1284$ & $0.2973$ 
& $14.4$
&  $21.0$ 
& \multirow{2}{*}{$0.0397$}
\\
\hhline{|~|-|-|-|-|-|-|-|~|}
 & $a_0$  & $2.49$ & $2.20$ & $0.1191$ & $0.2960$  
 & $27.0$ 
 &  $7.33$ 
 &
 \\
 \hline\hline
  \multirow{2}{*}{GBW} 
& $K_0^*$ & $2.97$ & $2.30$ & $0.1297$ & $0.3070$ 
& $15.6$ 
&  $28.5$ 
& \multirow{2}{*}{$0.0539$}
\\
\hhline{|~|-|-|-|-|-|-|-|~|}
& $a_0$  & $2.53$ & $2.24$ & $0.1201$ & $0.2978$  
 & $29.1$ 
 &  $8.26$  
 &
 \\
 \hline\hline
  \multirow{2}{*}{EDBW} 
& $K_0^*$ & $3.20$ & $2.34$ & $0.1339$ & $0.3091$ 
& $6.10$
&  $14.2$ 
& \multirow{2}{*}{$0.0182$}
\\
\hhline{|~|-|-|-|-|-|-|-|~|}
& $a_0$  & $3.08$ & $2.65$ & $0.1318$ & $0.3156$  
 & $13.3$ 
 &  $6.56$ 
 &
 \\
 \hline\hline
  \multirow{2}{*}{EGBW} 
& $K_0^*$ & $3.31$ & $2.49$ & $0.1358$ & $0.3163$ 
& $5.28$ 
&  $18.3$ 
& \multirow{2}{*}{$0.0119$}
\\
\hhline{|~|-|-|-|-|-|-|-|~|}
& $a_0$  & $3.14$ & $2.69$ & $0.1330$ & $0.3173$  
 & $13.2$ 
 &  $7.04$  
 &
 \\
 \hline
\end{tabular}
\caption{
Optimized and fitted values for the continuum thresholds and scale factors  are given for each model and channel, along with the (dimensionless) values  of $\{\chi_\Lambda^2,\chi_{\Lambda'}^2\}$ and $\Delta$ that quantify the scale factor properties of energy-independence and measure of universality.  All units in GeV, except for the dimensionless quantities $\chi^2$ and $\Delta$.
Resonance models follow the Table~\ref{tab:my_label} ordering of increasing sophistication and  increasing width effect.}
\label{scale_factor_table}
\end{table}

The narrow resonance model leads to non-physical values of the continuum that are well below the  $K_0^*(1430)$ mass and somewhat below the $a_0(1450)$ mass, so the Gaussian QCD sum-rule analysis rules out the NR (narrow resonance) model, a result which is not particularly surprising given the broad nature of the $K_0^*(700)$. This finding that emerges from the optimization procedure aligns with Fig.~\ref{kappa_model_fig} which illustrates that the Gaussian sum-rules can clearly distinguish the NR model from all other models that include width effects. 

The PROP model results in Table~\ref{scale_factor_table} can also be compared with the benchmark analysis of Ref.~\cite{CLQCDSR_2020}, which allows us to explore the effect of the revised scale factor methodology outlined in Section~\ref{scale_factor_section}.  In comparison to Ref.~\cite{CLQCDSR_2020}, Table~\ref{scale_factor_table} shows an approximate $10\,{\rm MeV}$  increase in all scale factors, a similar value of $\Delta$, and a general tendency of increased values of the smallest continuum value in each channel. This demonstrates that the revised methodology, which avoids use of the off-diagonal constraint \eqref{Pi_12_constraint}, leads to a change in the scale factors similar to the range seen across the various models  in Table~\ref{scale_factor_table}. 
This indicates that the  off-diagonal constraint \eqref{Pi_12_constraint} should be considered as an effect  similar to the resonance model dependence.
However, unlike the benchmark methodology which associates a continuum value with each resonance, the revised methodology [see Eqs.~\eqref{GSR_11} and \eqref{GSR_22}]  
associates the continuum with a mixture of the two states, 
so the  $s_0^{(2)}$ value slightly below the $K_0(1430)$ mass scale in the $\kappa$ case  suggests that refinement of the PROP model is necessary.

Important trends in the scale factors, measures of scale-factor energy-independence $\chi^2$, and measure of scale-factor universality $\Delta$ can be seen in Table~\ref{scale_factor_table} and in Figs.~\ref{isodoublet_scale_factor_models_fig}--\ref{isotriplet_scale_factor_fits_fig}.  Fig.~\ref{isodoublet_scale_factor_models_fig} and Fig.~\ref{isotriplet_scale_factor_models_fig} show that the scale factors increase as the model sophistication/width  increases  (see  Table~\ref{tab:my_label} for sophistication/width hierarchy).  There is also a trend of increasing continuum with increasing model sophistication/width, leading to progressively better separation of the continuum from  the resonance mass scales, providing the desired necessary improvements to the PROP model. 

Fig.~\ref{isodoublet_scale_factor_fits_fig} and Fig.~\ref{isotriplet_scale_factor_fits_fig} illustrate that the extended models (EDBW, EGBW) provide a better match between the theoretical prediction and hadronic model by showing less deviation from  a constant scale-factor value as a function of $\hat s$. This behaviour  is quantified in  Table~\ref{scale_factor_table} by the clear decrease in 
$\{\chi^2_\Lambda,\chi^2_{\Lambda'}\} $ for the extended models (EDBW, EGBW) compared to the non-extended models (DBW, GBW). The small deviation of the scale factors from a constant value provides evidence for the scale factor connection between QCD operators and Chiral Lagrangian fields as expressed by Eqs.~(\ref{M_scale_new},\ref{scale_factors_new}), and that no significant additional dynamical elements are necessary for this connection.   

The most important feature of Table~\ref{scale_factor_table} is the distinct improvement in the $\Delta$ measure of scale factor universality in the EDBW and EGBW extended models.  This  demonstrates that the scale factor Gaussian QCD sum-rule methodology has sufficient sensitivity to distinguish between different hadronic resonance models. Remarkably, the extended EDBW and EGBW  models  that  emerge from  fits to
$\pi K$ scattering data and $\pi\eta$ scattering  calculations (see Section \ref{res_shape_section}) lead to the closest alignment with scale-factor universality. Thus the scale factor bridge between Chiral Lagrangians and QCD sum-rules provides a sensitive probe of the connection between quark-level QCD predictions and low-energy hadronic physics.

\section{Summary}
\label{summary_section}

In this paper we have explored and developed several aspects of the  relationship between low-energy hadronic physics, Chiral Lagrangians, and QCD sum-rules. In Section~\ref{res_shape_section}
a new background-resonance interference approximation was developed to describe the $\pi K$ and $\pi\eta$ scattering amplitudes.  In this approximation inspired by Chiral Lagrangians, $\pi K$ scattering data were fitted by a background-resonance interference approximation as first expressed in Eq.~\eqref{E_T012_rBW} and then extended to an energy-dependent unitarized from in Eq.~\eqref{E_T012_rEBW}.  Similarly, for $\pi\eta$ scattering (where there is no data available), theoretical calculations of the scattering amplitude were fitted to Eq.
\eqref{E_T01_rBW} and the local unitarized version in Eq.~\eqref{E_T01_a0_EBW}.  The  background-resonance interference approximation provides an excellent description of the both $\pi K$ and $\pi\eta$ scattering as illustrated in Figs.~\ref{F_rBW}--\ref{F_rEBW} and Figs.~\ref{F_pieta_fit_BW}--\ref{F_pieta_fit_EBW}.

Scale factor matrices were introduced in  Eq.~\eqref{M_scale_new} to relate Chiral Lagrangian mesonic fields to QCD (two-quark and four-quark) composite operators.  Chiral symmetry was used to constrain the scale factor matrices in Eq.~\eqref{scale_factors_new} so that they contain only two scale factor parameters 
$\{\Lambda,\Lambda'\}$ that are universal across the scalar and pseudoscalar multiplets.  A revised Gaussian QCD sum-rule methodology that enables the extension to higher-dimensional multiplets (e.g., isoscalars including mixing with a glueball component) was outlined in  Section~\ref{new_scale_methodology_sec}.  The key feature of the revised methodology is circumventing the off-diagonal constraint \eqref{Pi_12_constraint} for two-dimensional isospin subsystems and similar constraints that would arise in higher-dimensional subsystems. Eqs.~(\ref{GSR_11},\ref{GSR_22}) summarize the new Gaussian QCD sum-rule methodology for the isodoublet/isotriplet sectors.    

An essential ingredient in the determination of the scale factors is the  Gaussian sum-rule of the resonance contributions  defined in Eq.~\eqref{gsr_res}.  These resonance contributions get combined with QCD contributions and other input parameters for determining the scale factors via Eqs.~\eqref{scale_relation_lambda} and \eqref{scale_relation_lambda_prime}.  
In Section~\ref{models_section} an increasingly sophisticated series of resonance models were developed (see Table~\ref{tab:my_label} for a summary), with the most sophisticated models emerging from the background-resonance interference approximation fits to $\pi K$ and $\pi\eta$ scattering.  Figs.~\ref{kappa_model_fig}--\ref{isotriplet_alt_solution_fig} illustrate that Gaussian sum-rules have sufficient resolution to distinguish between  different models. 

The impact of different resonance models on the predicted scale factors is examined in Section~\ref{analysis_section}, and the scale factors are assessed against the important properties of universality and energy-independent (constant) behaviour as a function of the Gaussian sum-rule energy scale  $\hat s$.  Table~\ref{scale_factor_table} summarizes our results for the isodoublet and isotriplet sectors for different resonance models.  Compared to the benchmark analysis of Ref.~\cite{CLQCDSR_2020}, the revised methodology has an impact on the scale factors similar to the range across different models in Table~\ref{scale_factor_table}, indicating that the off-diagonal constraint \eqref{Pi_12_constraint} has an impact similar to the role of resonance models. 
The EDBW and EGBW models that emerge from the background-resonance interference approximation fits to $\pi K$ and $\pi\eta$ scattering show a distinct improvement in the measures of universality and energy-independence as quantified by  
$\{\Delta,\chi^2_\Lambda,\chi^2_{\Lambda'}\}$ in
Table~\ref{scale_factor_table}.

In conclusion, 
a revised methodology has been developed for the connection between Gaussian QCD sum-rules and Chiral Lagrangians. The key advantage of this revised methodology is that it provides a framework that can be naturally extended to higher-dimensional isospin sectors, including mixing with glueballs.    
Furthermore,  a new
background-resonance interference approximation has  been developed
that provides an excellent description of $\pi K$ and $\pi\eta$ scattering.    This  background-resonance interference approximation inspires new  isodoublet and isotriplet  resonance models that can be used within a Gaussian sum-rule analysis of these sectors.  The scale factors connecting Chiral Lagrangian mesonic fields to QCD have been analyzed for the isodoublet and isotriplet sectors using a series of increasing sophisticated resonance models, culminating in the    models inspired by the background-resonance interference approximation.  Remarkably, the background-resonance interference approximation models lead to scale factors that best exhibit the crucial properties of universality and energy independence.  
Thus the scale factor bridge between Chiral Lagrangians and QCD sum-rules provides a sensitive probe of the connection between quark-level QCD predictions and low-energy hadronic physics.  With these new methodologies firmly established,  a clear framework is in place for future studies of the  challenging   mixings of the scalar isoscalar sector, including glueball components.

\section*{Acknowledgments}
TGS acknowledges  research funding from the Natural Sciences and Engineering Research Council of Canada (NSERC).  TGS is grateful to SUNY Polytechnic Institute for hosting a sabbatical visit that initiated this work.

\end{document}